\documentclass[%
 reprint,
 amsmath,amssymb,
 aps,
pra,
]{revtex4-2}
\usepackage{soul}
\usepackage{graphicx}
\usepackage{dcolumn}
\usepackage{bm}
\usepackage{hyperref}
\usepackage{xcolor}

\usepackage{float}
\usepackage{fancybox}

\usepackage{setspace}
\usepackage{quantikz}

\newcommand\numberthis{\addtocounter{equation}{1}\tag{\theequation}}

\newcommand{\vech}[1]{ {\mathop{#1} \limits^{ \vbox to -1pt{\kern-1.5pt\hbox{\kern0.5pt$\scriptscriptstyle\rightharpoonup$\kern-0.5pt}} } } }

\newlength{\ketketwidth}
\newlength{\ketwidth}

\newcommand{\ketet}{\rangle\kern-0.2em\rangle}
\newcommand{\brara}{\langle\kern-0.2em\langle}

\newcommand\froogle{\raisebox{3.5pt}{\scalebox{0.7}[0.7]{$\bm{\square}$}} \kern-1.4pt\raisebox{-1.0pt}{\scalebox{0.7}[0.7]{$\bm{\square}$}}}

\begin{document}

\preprint{APS/123-QED}

\title{
Quantum simulation of gauge theories on dynamical spacetimes \\ via Floquet-induced matrix models
}

\author{Samuel Buckley-Bonanno, Noah Eckstein, Susanne F. Yelin}
\affiliation{
Department~of~Physics,~Harvard~University,~Cambridge,~MA~02138,~USA
}

\date{
  July 4, 2026 
}

\begin{abstract}
  Quantum simulations of gauge theories are typically built on spatial lattices, an approach that has enabled major progress at the cost of requiring fixed background geometries and obscuring the treatment of curved and dynamical spacetimes. 
  Large-$\textit{N}$ matrix models offer an alternative, encoding spacetime geometry and gauge fields in the commutation structure of a set of Hermitian matrices, with the classical continuum emerging smoothly at large matrix dimensions. Here we introduce a Floquet framework that makes these models directly accessible to programmable quantum platforms. 
  We show that Euclidean path integral weights of a Yang-Mills matrix models are reproduced, at leading order in the coupling, by the ensemble-averaged fidelities of Haar-random states evolved under periodic sequences of matrix operators. The observables for the simulated matrix model can then be accessed through established randomized benchmarking protocols in terms of the Loschmidt echo. 
  The encoding requires exponentially fewer qubits than canonically quantized approaches. Numerically, we validate the fidelity-weight correspondence, demonstrate parallelized quantum circuits that sample the path-integral measure, and identify the deconfinement transition of an $\textit{SU}\,\hspace{-0.04cm}\text{(2)}$ gauge field on both flat and expanding cosmological backgrounds. 
  By avoiding a fixed spacetime lattice, the framework preserves continuous symmetries and unitarity on dynamical geometries, opening quantum simulation to field and spacetime dynamics beyond the reach of conventional lattice methods.
\end{abstract}

\maketitle

\section*{Introduction}

\begin{figure*}
  \includegraphics[width=\textwidth]{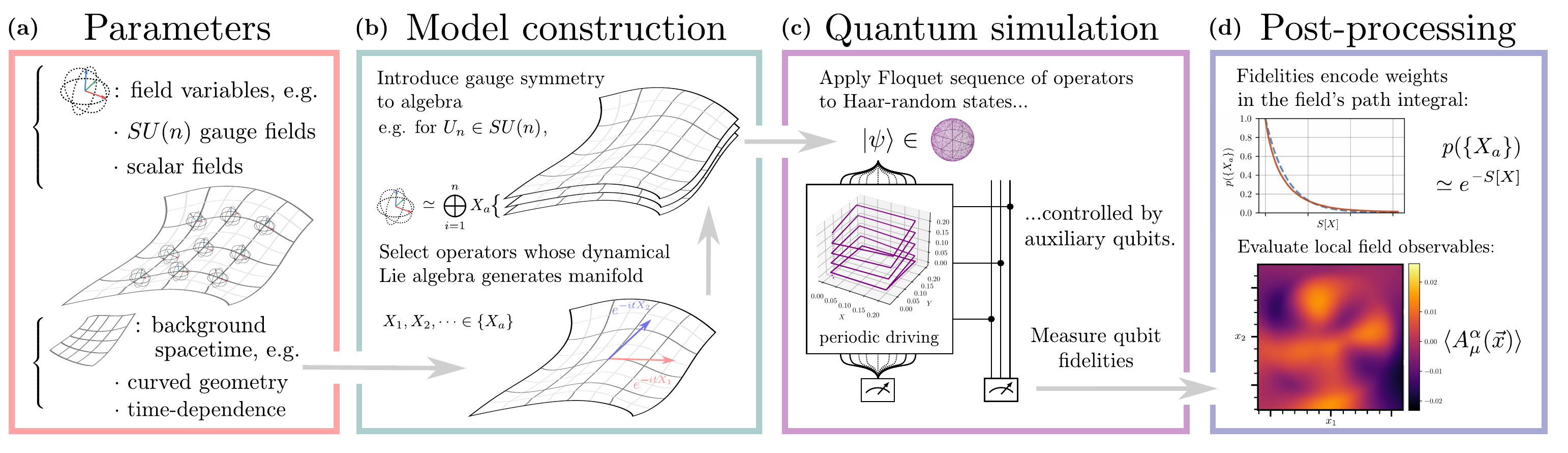}
  \caption{\label{fig:flowchart}
  \textbf{Matrix model quantum simulation procedure}. \textbf{(a)} Given a specification of bosonic fields on a spacetime, there is an associated matrix state that encodes their Lagrangian. The fields can be scalar or Yang-Mills or other gauge fields, and the background spacetime can be curved and time-dependent. \textbf{(b)} The construction of the associated matrix state proceeds from these parameters. The spacetime geometry is encoded first by selecting a set of Hermitian operators whose dynamical Lie algebra maps in the commutative limit to the desired manifold. Field symmetries are then introduced to the set of matrices: $U(n)$ non-commutative Yang-Mills fields are formed with a direct sum of $n$ copies of the original matrix state, while scalar fields are added via an additional matrix. \textbf{(c)} To study the matrix-model path integral, the Hermitian operators that make up a given matrix state are applied to a Haar-random ensemble of states as a periodic Floquet sequence. The average fidelity after this sequence is proportional to the weighting of the matrix state in a Euclidean matrix path integral. Multiple Floquet sequences can be applied in parallel via operations controlled by auxiliary control qubits, allowing sampling across an ensemble of matrix states. \textbf{(d)} Once the weighting of the matrix states has been measured, field and spacetime geometry information can be extracted by post-processing the matrix states. The fidelities map onto the partition function weights $e^{-S[X]}$, and the field observables $ \langle A_\mu^a(\vech{x}) \rangle $ and $\langle F_{\mu\nu}^a(\vech{x})\rangle$ can be reconstructed and localized using quasi-coherent states $|\vech{x}\rangle$, which pick out optimally localized expectation values of the matrix algebra for a given classical spacetime coordinate $\vech{x}$.}
\end{figure*}

A fundamental challenge for the quantum simulation of gauge theories is the conflict between continuous spacetime symmetries and the discrete resources of conventional computation. Lattice gauge theory resolves this by fixing a spatial grid, trading exact translational and rotational symmetry for computational tractability. This strategy is remarkably successful for static background geometries but becomes problematic whenever spacetime is itself dynamical, as with cosmological expansion or near gravitational horizons \cite{Yamamoto2014,Jersak1996,Campos1998}. Representing spacetimes such as these with lattices requires adding or removing lattice sites, resulting in violations of unitarity that are difficult to repair \cite{Cotler2022}. Most proposals for quantum simulations of gauge theories \cite{Byrnes2005,Preskill2012,Meth2025,Gonzalez-Cuadra2025} retain the lattice framework, partly because many quantum platforms are themselves arranged as physical lattices. A lattice discretization, however, is fundamentally a classical treatment of position regardless of whether the on-site states are quantum.

Quantum processors offer an alternative, with their associated unitary operators and continuous Hilbert spaces already carrying their own continuous symmetries. We make use of this by drawing on the theory of large-$N$ Yang-Mills matrix models \cite{tHooft1974,Eguchi1982,Gonzalez-Arroyo1983_1}, in which spacetime geometry is encoded in the commutation structure of a set of $N\times N$ Hermitian matrices $\{X_a\}$ rather than in a lattice of points. The classical continuum emerges smoothly as $N\to\infty$, while for finite $N$ every physical quantity is still represented continuously on a non-commutative manifold and extracted via expectation values of quantum operators. Gauge symmetries appear as literal unitary symmetries of the matrix operators, and dynamical spacetimes, including cosmological metrics, are encoded by adjusting the commutation structure of $\{ X_a \}$, without adding or removing degrees of freedom.

Our central result is a quantum simulation protocol for large-$N$ Yang-Mills matrix models where the Euclidean path-integral weights of the matrix model can be extracted from the average fidelities of Haar-random quantum states undergoing periodic Floquet sequences. This protocol relies on a framework of matrix models \cite{Steinacker2010,Berenstein2012,Steinacker2024} in which a set of $N\times N$ Hermitian matrices ${X_a}$ carries both the spacetime and field content for the simulated state. In this framework, the role of derivatives in the equations of motion is played by commutators: expressions of the form $[X_a,\cdot]$ acting on arbitrary operators serve as non-commutative derivatives, nested commutators such as $\sum_a [X_a,[X_a,\cdot]]$ define the analogue of a Laplace operator, and commutators $[X_a,X_b]$ constructed from the operators $\{X_a\}$ encode the local Poisson structure, curvature, and effective metric of the emergent geometry. Gauge and matter fields are introduced as additional matrix structure on top of this background: scalar fields appear as extra matrices appended to ${X_a}$, while Yang--Mills fields arise as fluctuations around block-diagonal matrix backgrounds with internal unitary symmetry.  Local classical information is extracted using quasi-coherent states $|\vec{x}\rangle$, which are the states whose expectation values for operators $\{X_a\}$ are most sharply localized around classical coordinates $\vec{x}$. Expectation values of these states give local field observables, providing access to local physical information in the matrix encoding at the cost of uncertainty in the result. This includes a tunable quantum uncertainty in the underlying continuum geometry, as governed by the non-commutative structure of the matrix states. 

This same commutation structure also determines the Euclidean matrix-model action $S[X]$, which is built from squared commutator terms of the form $\text{tr}([X_a, X_b]^2)$. 
Our primary observation is that these quantities can be accessed directly with a Floquet protocol.  Specifically, for each pair of matrix operators, a four-step, symmetric sequence $\mathcal{F}_2(X_a,X_b;\Delta t)$ has an effective Hamiltonian proportional to the commutator $i[X_a,X_b]$ at leading order in the Floquet frequency.  When this sequence is applied repeatedly to a Haar-random ensemble of states, the averaged fidelity, or Loschmidt echo, reproduces the Euclidean Boltzmann factor $e^{-S[X]}$.  Each fidelity measurement is therefore a direct probe of the matrix configuration's path-integral weight, and hence of the partition function of the matrix model.

On a quantum computer or programmable quantum simulator, the $N\times N$ matrices are encoded as Hamiltonian operators acting on $2\log_2 N$ qubits \cite{Rinaldi2022, Gharibyan2021}.  The quantum device performs the calculation by applying periodic sequences of these Hamiltonians to Haar-random states and reading out the resulting Loschmidt echoes, which can be accessed through standard randomized benchmarking protocols.  Controlled Floquet sequences then allow a register of control qubits to sample an ensemble of matrix states with probabilities proportional to their Euclidean weights.  Once these weights have been measured, geometric and field observables are extracted by post-processing the matrix states with quasi-coherent states.  The observables accessible through this protocol include partition-function weights, gauge-field expectation values on curved spacetime slices, Wilson-loop expectation values quantifying confinement and deconfinement, and operator product expansions of local observables in the matrix model.  Using these observables, we numerically demonstrate the deconfinement phase transition of an $SU(2)$ gauge field on both flat space and an expanding cosmological background, as an illustration of the capabilities of the protocol.  An overview of the quantum simulation protocol as a whole is provided in Fig.~\ref{fig:flowchart}.

\section{Matrix models and non-commutative geometry}

Originating in the study of the large-$N$ limit of quantum chromodynamics \cite{tHooft1974,Eguchi1982,Gonzalez-Arroyo1983_1,Gonzalez-Arroyo1983_2}, matrix models have inspired work ranging from representations of the quantum Hall effect \cite{Bellisard1994,Marcolli2006,Pasquier2007} to the foundations of gauge-string dualities \cite{Gubser1998,Polyakov2000} and the AdS/CFT conjecture \cite{Maldacena1997,Polyakov2012}. Leading candidates for $M$-theory, notably the BFSS \cite{BFSS1997,Seiberg1997} and IKKT models \cite{IKKT1997,IKKT1999}, are themselves matrix models. For our purposes, we focus on Yang-Mills matrix models, and take Steinacker's formalism \cite{Steinacker2010,Steinacker2024} for covariant matrix states \cite{Snyder1947,Sperling2019,Steinacker2017} and quasi-coherent states \cite{Berenstein2012,Steinacker2021} as the centerpiece of our framework. With these tools, gauge theories are expressed through matrix commutators rather than spatial derivatives, with the algebraic structure of derivatives arising via the Jacobi identity taking the place of the Leibniz product rule.

The matrix model construction begins from the Yang-Mills action written in terms of covariant derivatives, $S_\text{YM} = -\frac{1}{2g^2} \int d^Dx\, \text{Tr}\left(  [D_\mu, D_\nu][D^\mu, D^\nu] \right)$, in which the field strength tensor appears as a commutator, $-i g T^a F_{\mu\nu}^a = i [D_\mu, D_\nu]$, measuring the curvature of a gauge bundle. Replacing the covariant derivatives $D_\mu$ with $N \times N$ Hermitian matrices $X_a$, and spacetime integration with a matrix trace, yields the large-$N$ Yang-Mills matrix model action,
\begin{equation} \label{eq:ym_action}
S[X] = \frac{1}{g^2} \text{Tr} ([X_a, X_b] [X^{a}, X^{b}]),
\end{equation}
with indices contracted by the Minkowski metric $\eta^{ab}$ for a Lorentzian model or the Euclidean metric $\delta^{ab}$ for the Euclidean model, the two being related by a generalized Wick rotation \cite{Nishimura2022}. The variation of $S[X]$ yields the equations of motion $\square X_a \equiv \sum_b [X_b, [X^b, X_a]] = 0$, where the matrix Laplacian $\square$ evaluates the quadratic Casimir of the algebra generated by the matrix state $\{X_a\}$ in the adjoint representation, and serves as the non-commutative analogue of the Laplace-Beltrami operator.

\begin{figure}
  \includegraphics[width=\linewidth]{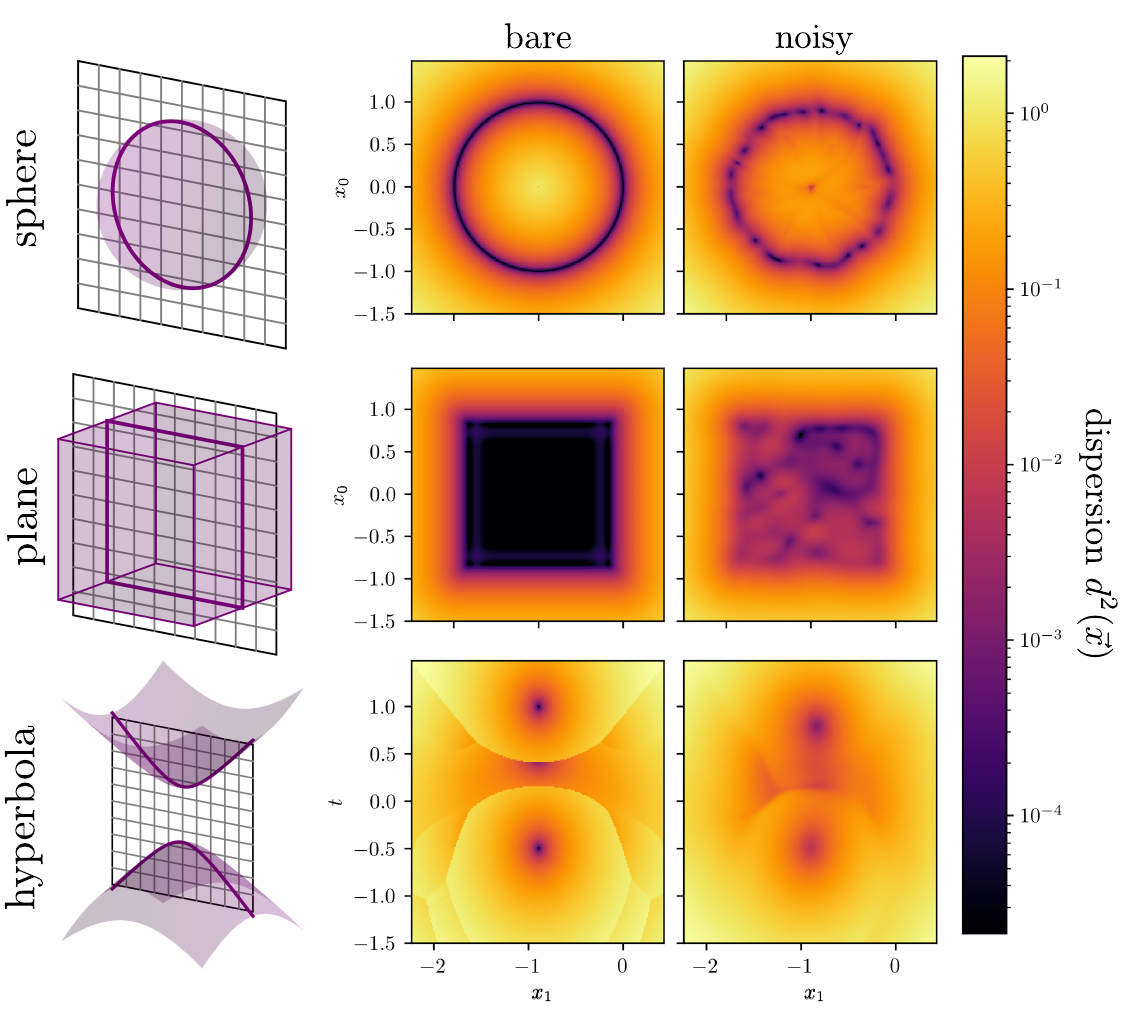}
  \caption{\label{fig:hexaplot}
  \textbf{Displacement plots of non-commutative geometries.} Elementary manifolds can be encoded as non-commutative geometries in terms of matrix representations. We show cross-sections for spherical, flat and hyperbolic geometries. For the non-commutative sphere, we use high-dimensional representations of $SU(2)$ angular momentum operators; the non-commutative plane is constructed from finite matrix representations of the canonical position $\hat{x}$ and momentum $\hat{p}$ operators via a Jordan-Schwinger construction; and Lorentzian spacetimes with hyperbolic geometry are encoded using operators drawn from the $SO(4,2)$ Lie algebra. The middle and right columns depict the displacements $d^2(\vec{x}) = \sum_i\left(\langle \vech{x}| X^i |\vech{x}\rangle-x^i\right)^2$ across two axes of variation $x_0$ and $x_1$, where $|\vech{x}\rangle$ are quasi-coherent states localized around a position vector $\vech{x} = \{ x_0, x_1, \dots\}$. Regions of low displacement mark where a good classical manifold emerges from the matrix state. The middle column shows the unperturbed matrix states, while the right column shows the same matrices perturbed by element-wise Gaussian noise with $\sigma = 0.02$, as a fraction of the maximal eigenvalue. As noise is added, these low-displacement regions are visibly degraded, indicating the breakdown of the classical limit.}
\end{figure}

As the matrix Laplacian suggests, the geometry implicit in a set of matrices is determined by their dynamical Lie algebra, i.e.~the algebra generated by $\{X_a\}$ under repeated commutation, whose coadjoint orbits cover a manifold that defines the effective classical background \cite{Kirillov2004,Hall2013}. Three geometrically distinct examples are illustrated in Fig.~\ref{fig:hexaplot}. For $\{ X_a \} = \{ r J_a^{(N)} \}$, with $J_a^{(N)}$ the $N$-dimensional representations of $SU(2)$ angular momentum generators, the relation $(r J_1)^2 + (r J_2)^2 + (r J_3)^2 = r^2 \openone$ mimics the constraint in Cartesian coordinates $x^2 + y^2 + z^2 = r^2$, indicating a non-commutative sphere of radius $r$ \cite{Madore1991,Hoppe1982,Grosse1996}. Canonically commuting operators $\{Q_i, P_i\}$ encode the flat Moyal-Weyl quantum plane, with vanishing matrix Laplacian. Minkowski space is represented by the generators $X^a = M^{a5}$ of $SO(4,2)$, whose coadjoint orbit obeys $-(X^0)^2 + \sum_{i=1}^4 (X^i)^2 = -R^2 \openone$, mimicking the Minkowski signature; scaling the space-like matrices relative to the time-like matrix yields a finite representation of the Friedmann-Lema\^itre-Robertson-Walker (FLRW) metric for cosmological simulations \cite{Steinacker2024}. Explicit finite constructions of all of these geometries are provided in Supplementary Information (S.I.)~\ref{app:jordan-schwinger}.

Quantum fields are introduced on these manifolds in two ways. Scalar fields are added as additional matrices $\phi_i$ appended to the set $\{X_a\}$. With the geometric matrices held fixed, the cross-commutator terms act as inner derivations on the coadjoint orbit, producing the action of a scalar field on a curved manifold with quartic self-interactions. Yang-Mills gauge fields with $SU(n)$ symmetry are introduced by taking a direct sum of $n$ copies of the background matrices, $\bar{T}_a = \bigoplus^n X_a$, and perturbing it by $SU(n)$-symmetric fluctuations $T^a = \bar{T}^a + \mathcal{A}^a$; the effective action for fixed $\bar{T}^a$ is then the non-commutative Yang-Mills action $S_\text{eff}^\text{YM} = -\frac{1}{g^2} \text{Tr}\,F_{ab} F^{ab}$, with indices contracted by the curved metric $g_{ab}$ induced by $\bar{T}^a$. The explicit field-strength variables, gauge transformations and field expectation value formulas are collected in Appendices \ref{app:fields-matrix-geometries-a} and \ref{quasicoherent-b}. Wilson loops are directly computable from the matrix states by looping the coordinates for calculating expectation values along a closed spatial contour. These provide an order parameter for the confinement-deconfinement transition of the encoded gauge fields and are among the primary targets of the quantum simulation protocol.

To make these matrix models phenomenologically useful, a dictionary between non-commutative geometries and classical Riemannian manifolds is needed. Just as coherent states are the maximally classical states of a harmonic oscillator, there exist optimally localized quasi-coherent states $|\vech{x}\rangle$ for extracting classical expectation values from a matrix state $\{X_a\}$, each defined as the ground state of the quadratic Hamiltonian
\begin{equation} \label{eq:quasi-hamiltonian}
  H_{\vech{x}} = \sum_{a=1}^d \left( X_a - x_a \openone \right)^2,
\end{equation}
where $x_a$ is the $a$th component of a real-valued vector $\vech{x}$. As a result, the encoded information in a matrix state $\{X_a\}$ is as continuous as the space of quasi-coherent states, generated for any $\vech{x} \in \mathbb{R}^d$. Each classical point carries a finite dispersion $\Delta^2(\vech{x})$ and displacement $d^2(\vech{x})$, quantifying the quantum uncertainty at $\vech{x}$ and the error against the intended classical coordinates. The subset of $\vech{x}$ where both are minimal constitutes the classical manifold encoded by the matrices, with an induced metric recoverable from quasi-coherent expectation values of commutators as detailed in Appendix \ref{quasicoherent-b}. This formalism is a classical post-processing step, as the quantum simulation itself concerns only the dynamics of the matrix state $\{X_a\}$. With matrix model encodings, we trade lattice discreteness for quantum uncertainty in spatial information.

\section{Floquet simulation of the matrix path integral}

The matrix model Lagrangian can be quantized either by canonical quantization of each individual matrix entry as $(\hat{X}_a)_{ij}$, or with a path integral defined over the space of Hermitian matrices. After a Wick rotation, the Euclidean path integral for a matrix action $S[X]$ is
\begin{subequations}
  \begin{align}
    Z &= \int dX\,e^{-S[X]} \\
    \langle \mathcal{O} \rangle &= \frac{1}{Z}\int dX\,\mathcal{O}\,e^{-S[X]} \label{eq:ope},
  \end{align}
\end{subequations}
with Euclidean weights related to Lorentzian observables via analytic continuations of a specific form
for matrix models \cite{Rinaldi2022,Nishimura2022}. These two quantizations require very distinct quantum resources on quantum simulators. Previous proposals \cite{Rinaldi2022, Gharibyan2021} rely on a truncation of the Fock space of canonically-quantized operators $(\hat{X}_a)_{ij}$, requiring $\Lambda N^2 d$ qubits for $d$ distinct $N \times N$ matrices truncated to $\Lambda$ orthogonal states per entry, with severe multi-qubit gate overheads \cite{Barenco1995,Jandura2022,Evered2023,Saffman2016,Vidal2002,Bennett2002,Hong-Ye2025}. Canonical quantization also neglects the natural role of the Hermitian matrices $\{ X_a \}$ as quantum operations in themselves, acting on a space of quasi-coherent states \cite{Starodubtsev2003,Steinacker2021,Steinacker2024}. We therefore encode the $N \times N$ matrices directly as Hamiltonian operators acting on $2 \log_2 N$ qubits. Treating the matrices as operators rather than states makes a fidelity-based quantum simulation of the matrix path integral tractable.

\begin{figure*}
  \includegraphics[width=\textwidth]{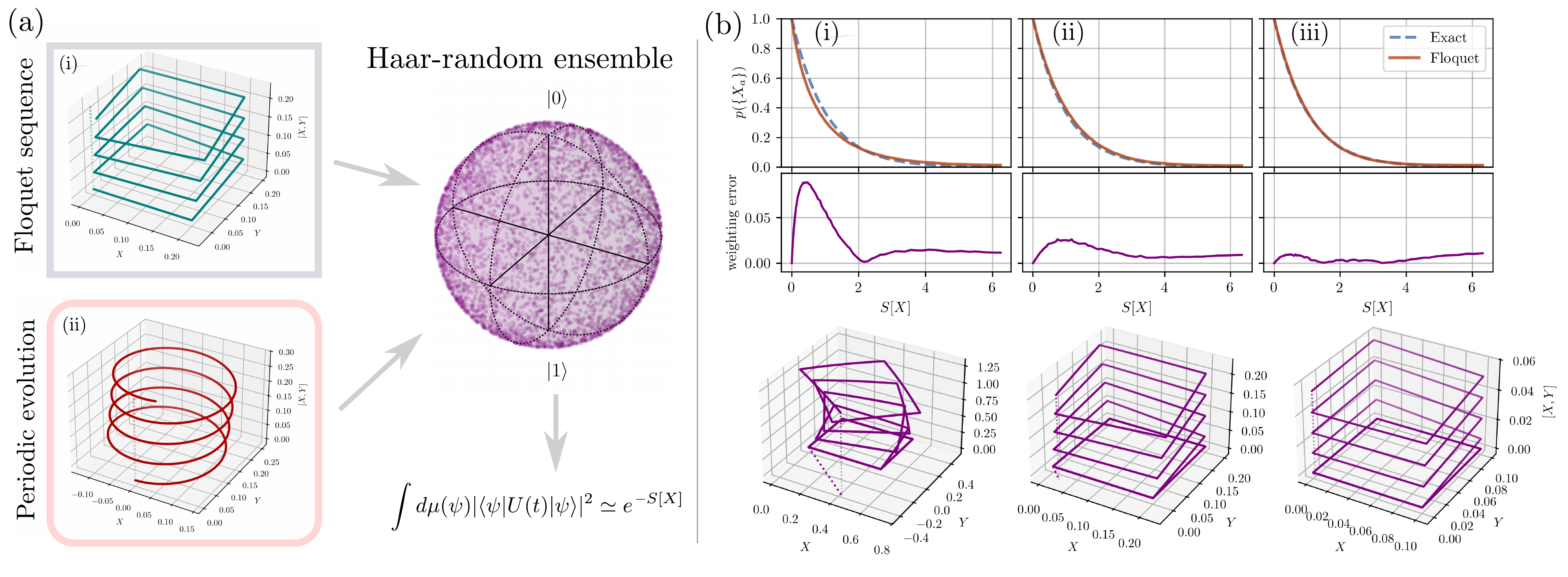}
  \caption{\label{fig:fidelities}
    \textbf{Comparison of fidelities and the matrix model partition function.}
\textbf{(a)} For a single matrix state $\{X_a\}$, a periodic Floquet sequence is applied over the matrix operators. Plots (i) and (ii) show the resulting unitary evolution in operator space for Trotterized and uniform periodic driving, respectively. The axes denote coefficients in the effective Hamiltonian expansion during the evolution. The leading-order terms are commutators of the alternating operators $X_a$ and $X_b$. When applied to a Haar-random ensemble of states, the fidelity fall-off in the high-frequency limit scales as $\exp(-\mathrm{tr}(H^2))$, and for sequences over multiple operator pairs reproduces the Euclidean matrix-model action $S[\{X_a\}]$.
\textbf{(b)} Average fidelities $\langle f\rangle$ of Haar-random states under Floquet evolution $\mathcal{F}_2(\{X_a\}; \Delta t )$ are compared with the expected Euclidean weights $e^{-S[\{X_a\}]}$ for an ensemble of matrix states. Columns (i)--(iii) show results for increasing Floquet frequencies $\hbar\omega_F/|\lambda_N| = 3.3\times10^4$, $1.0\times10^5$, and $3.3\times10^6$. The operators are $100\times100$ Hermitian matrices encoding the Moyal-Weyl quantum plane, perturbed by element-wise Gaussian noise over 50 variance increments up to $\sigma/|\lambda_N|=1.5$. For each increment, 100 perturbed matrix states are sampled, and each Floquet sequence is evaluated on 100 Haar-random states. As the Floquet frequency increases, the measured fidelities approach the exact Euclidean weighting $e^{-S[\{X_a\}]}$, while the remaining deviation is due to higher-order terms in the expansion of $\mathcal{F}_2$.}
\end{figure*}

In particular, we would like an observable whose outcome probabilities conform to the Boltzmann weights $e^{-S[X]}/Z$ of the Euclidean matrix path integral. Expanding to leading order in the coupling,
\begin{equation}\label{eq:frob-partition}
\frac{e^{-S[X]}}{Z}\;\simeq\;\frac{1}{Z}\!\left(1
-\frac{1}{g^2}\sum_{a,b}\|i[X_a,X_b]\|_F^2\right),
\end{equation}
the action is a sum of squared Frobenius norms $\|A\|_F^2=\mathrm{Tr}(A^\dagger A)$ of commutators. This structure points directly toward fidelity measurements on periodically-driven quantum states, for which the unitary evolution is generated by commutators of Hamiltonians.

Let $X_a$ and $X_b$ be Hermitian operators serving as Hamiltonians. The four-step symmetric sequence
\begin{align*}
\mathcal{F}_2(X_a,X_b;\Delta t )
&= e^{i\Delta t X_b}e^{i\Delta t X_a}e^{-i\Delta t X_b}e^{-i\Delta t X_a}
\\ &= \exp\!\left(-i\Delta t ^2(i[X_a,X_b])+\mathcal{O}(\Delta t ^3)\right) \numberthis \label{eq:F2}
\end{align*}
isolates the commutator $i[X_a,X_b]$ at second order in $\Delta t $ when expanded via the Baker-Campbell-Hausdorff formula \cite{Baker1905,Campbell1897,Hausdorff1906}, where $\Delta t$ is the Floquet timestep. All first-order terms cancel by construction, leaving a dynamical decoupling sequence whose residual ``error'' is precisely the commutator term whose norm we wish to measure. The path through operator space over one cycle is depicted in Fig.~\ref{fig:fidelities}a.

When any unitary $U$ acts on a state drawn uniformly from the Haar measure on $\mathbb{C}^N$, the expected fidelity can be evaluated via Weingarten calculus. As detailed in \cite{CollinsSniady2006,Horodecki1999,Nielsen2002} as well as S.I.~\ref{app:haarrelations}, the result is
\begin{equation}\label{eq:weingarten}
\langle f\rangle  = \int d\mu(\psi) |\langle \psi | U |\psi\rangle|^2 = \frac{N+|\mathrm{Tr}(U)|^2}{N(N+1)}.
\end{equation}
Applying this to $U=\mathcal{F}_2(X_a,X_b;\Delta t)$ and expanding $|\mathrm{Tr}(\mathcal{F}_2)|^2$ to leading order in $\Delta t$ yields
\begin{equation}\label{eq:frob-haar}
\langle f_{ab}\rangle\;\simeq\;1-\frac{2\Delta t^4}{N+1}\|i[X_a,X_b]\|_F^2.
\end{equation}
Comparing this result to the expression for the weights of the Euclidean path integral in Eq.~(\ref{eq:frob-partition}), we see that their leading-order behaviors are identical under the identification $ \frac{1}{g^2} = \frac{2\Delta t^4}{N+1} $. Trotterizing $\mathcal{F}_2$ over all pairs from a matrix state $\{X_a\}$, we can reconstruct the full matrix model action $S[X]$ across all matrix pairs, and thus obtain the central result:
\begin{equation}\label{eq:fid-relation}
e^{-S[X]}\;\simeq\;\int d\mu(\psi)\,
\left|\!\left\langle\psi\,\middle|\,\prod_{a,b}\mathcal{F}_2(X_a,X_b;\Delta t)\,\middle|\,\psi\right\rangle\!\right|^2.
\end{equation}
Every measurement of the ensemble-averaged fidelity of a Haar-random state under the composite Floquet sequence therefore directly evaluates the Euclidean Boltzmann weight for that matrix configuration. The approximation is controlled by $\Delta t^6\,\mathrm{tr}([X_a,[X_a,X_b]]^2)\ll 1$, satisfied in the high-frequency regime at large $\omega_F=2\pi/\Delta t$, or the weak-coupling regime at small $\|X_a\|$. The drive time $\Delta t$ and Hilbert space dimension $N$ together set the effective coupling. 
\vspace{-0.45cm}

The full sequence can be compiled with only $d$ rather than $d^2 - d$ operations for a $d$-element matrix state, with the use of phase offsets in the Floquet sequence. Evaluating $\langle f\rangle$ is equivalent to global $U(N)$ twirling followed by gate fidelity estimation, making the protocol directly compatible with established randomized benchmarking methods. 
Moreover, the same statistics emerge from any periodic Hamiltonian alternating between matrix operators, since the Magnus expansion of any periodic drive possesses a leading-order effective Hamiltonian composed of a sum of commutators, detailed in Supplementary Information ~\ref{app:periodic-drive}. The Loschmidt echo of a generic periodic drive thus also takes the form of a weight in a matrix model path integral.

$$  $$

\section{Parallelized quantum circuits}

\begin{figure*}
  \includegraphics[width=\textwidth]{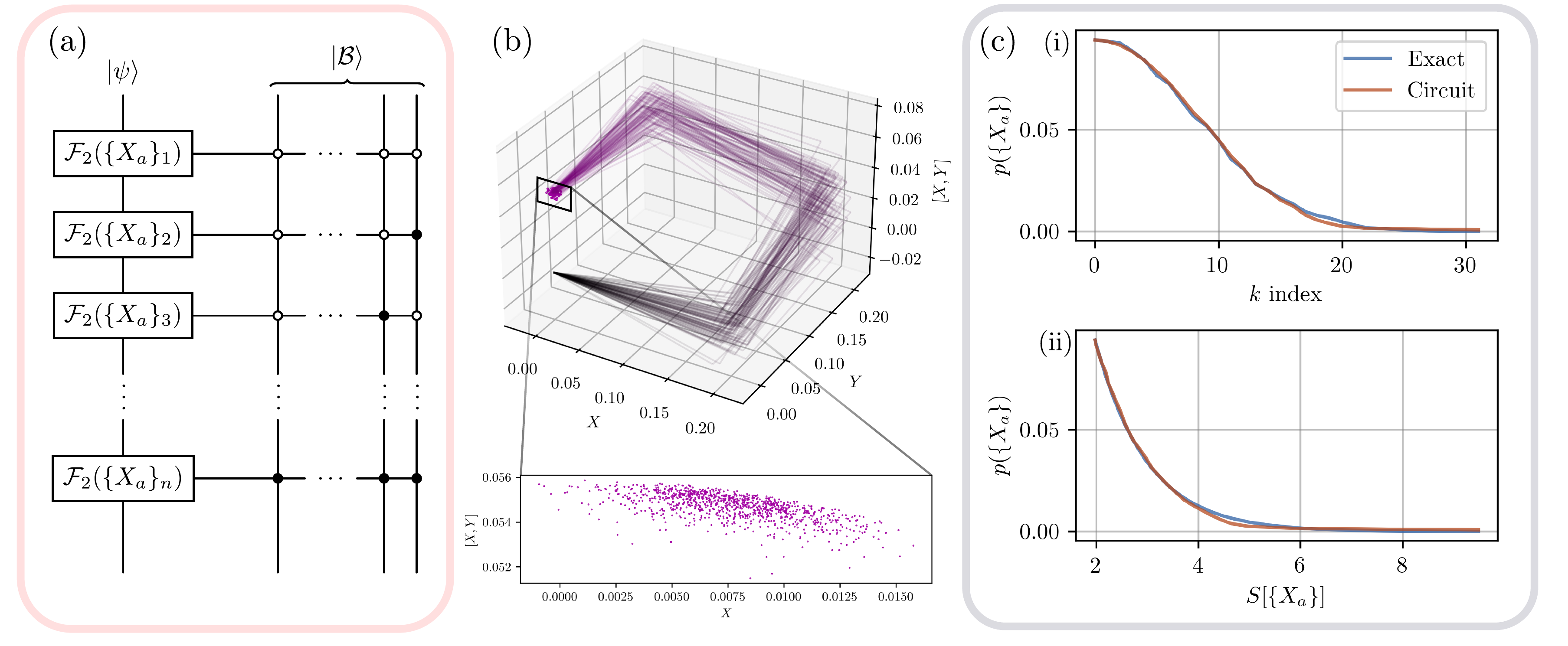}
  \caption{\label{fig:parallel-circuit}
    \textbf{Globally controlled quantum circuit and path-integral comparison.}
\textbf{(a)} Diagram of the parallel quantum circuit for sampling from the matrix path integral, as detailed in Appendix \ref{globally-controlled-d}. The Floquet sequence $\mathcal{F}_2$ for each matrix state $\{X_a\}$ in the ensemble is conditioned on a specific element of the entangled control-qubit basis $|\mathcal{B}\rangle$.
\textbf{(b)} Sample of 1000 paths through the space of effective Hamiltonians over one cycle of $\mathcal{F}_2(\{X_a\}_n; \Delta t )$ for a Gaussian distribution of perturbations. In this illustrative example, the perturbed matrix state is a $2\times2$ representation of the non-commutative sphere, given by the Pauli matrices. The distribution of path endpoints is shown, with the unperturbed endpoint marked in red. The perturbation variance is $\sigma/|\lambda_N|=0.01$, where $|\lambda_N|$ is the maximal eigenvalue of the unperturbed matrix state. The elongated distribution indicates greater variance in the operators than in their commutators, while the bias toward larger $[X_a,X_b]$ coefficients indicates a greater density of higher-action matrix states, as the majority of perturbed matrix states are perturbations away from a local minimum.
\textbf{(c)} Comparison of fidelities from a numerical simulation of the parallel quantum circuit with an ideal Monte-Carlo simulation of the Euclidean path integral over the same ensemble. The matrix state is a $100\times100$ Moyal-Weyl quantum plane, perturbed by element-wise Gaussian noise up to $\sigma/|\lambda_N|=0.02$. The Floquet sequences are applied at $\hbar\omega_F/|\lambda_N|=224$ over 1375 cycles. Using five control qubits, the circuit samples 32 distinct matrix states. After post-selection on stationary target-qubit outcomes, the probabilities of the control-basis states are shown in (i) as a function of the basis index $k$, alongside the exact Euclidean probabilities for the corresponding matrix states $\{X_a\}_k$. In (ii), the same probabilities are plotted against the expected action $S[\{X_a\}]$, showing that the circuit reproduces the exponential weighting $p(\{X_a\})\simeq e^{-S[\{X_a\}]}$ of the Euclidean matrix path integral.}
\end{figure*}

Using the Floquet sequence $\mathcal{F}_2$ applied to Haar-random states, the path integral weight of an individual matrix state is evaluated as an analog quantum operation. The quantities of interest for field simulations, however, are operator product expansions $\langle \mathcal{O} \rangle$ as given in Eq.~(\ref{eq:ope}), evaluated across an entire measure of matrix states at once. With controlled Floquet sequences and entangled states in a digital quantum circuit, $\langle \mathcal{O} \rangle$ can be evaluated in parallel, with the branches of the entangled state spanning an entire path integral measure, as presented in Fig.~\ref{fig:parallel-circuit}a. Each Floquet sequence $\mathcal{F}_2(\{X_a\}_k; \Delta t)$ is conditioned on a distinct computational basis state $|B_k\rangle$ of $n$ control qubits, which act as a binary index over $2^n$ matrix states. The control register is prepared in the even superposition
\begin{equation}
  |\mathcal{B}\rangle = \frac{1}{\sqrt{2^n}}\sum_k |B_k\rangle,
\end{equation}
so that $n$ control qubits suffice to activate $2^n$ distinct samples from the path integral measure. After the controlled sequences are applied to a Haar-random target state and the outcome is post-selected on the target remaining stationary, the relative frequencies of the control-qubit outcomes are weighted as
\begin{equation} \label{eq:pk}
  \frac{f_k}{\sum_k f_k} = \frac{e^{-S[\{X_a\}_k]}}{\sum_k e^{-S[\{X_a\}_k]}} = p_k,
\end{equation}
reproducing the probability distribution of the Euclidean path integral, from which any operator product expansion is evaluated as $\langle \mathcal{O} \rangle = \sum_k p_k \mathcal{O}_k$. The number of control qubits is logarithmic in the size of the discretization of the path integral measure. The full derivation of the circuit, the experimental preparation and post-selection of Haar-random states, and a locally-controlled variant that eliminates the multi-qubit Toffoli gate overhead by inserting individually-controlled perturbations into a global Floquet drive, are presented in Appendix \ref{locally-controlled-e}.

\section{Results and discussion}

To test the fidelity protocol, we numerically evaluate the expected fidelities $\langle f \rangle$ over Haar-random ensembles across Hermitian matrix variables with varying element-wise Gaussian noise, as shown in Fig.~\ref{fig:fidelities}. The error between the ideal path integral distribution and the measured fidelity diminishes at higher Floquet frequencies, as expected for a Trotterization procedure, arising primarily from higher-order BCH terms of order $\mathcal{O}(\Delta t^3)$. The error can be arbitrarily mitigated for high-probability states, but converges to a fixed floor for low-probability states, set by the minimal expected overlap $|\langle \psi |\phi\rangle|^2 \simeq 1/N$ of Haar-random states. Requiring the weights $e^{-S[X]}$ to lie above this floor bounds the accurately sampled actions as $S[X] \lesssim \log N$. Since $N$ scales exponentially with qubit number, this bound is linear in the number of qubits.

Figure~\ref{fig:parallel-circuit} demonstrates the globally controlled circuit architecture for evaluating operator product expansions $\langle\mathcal{O}\rangle=\sum_k p_k\mathcal{O}_k$ across a discrete ensemble of matrix states, encoding $2^n$ matrix states with only $n$ control qubits, while Extended Data Fig.~\ref{fig:scrambling-circuit} shows the locally controlled architecture in which perturbations to a fixed background matrix state are each activated by individual qubits, eliminating the multi-qubit controlled gate overhead. Both circuits reproduce the exponential weighting $p(\{X_a\})\propto e^{-S[\{X_a\}]}$ of the Euclidean path integral after post-selection on stationary target-qubit outcomes.

\begin{figure*}
  \includegraphics[width=\textwidth]{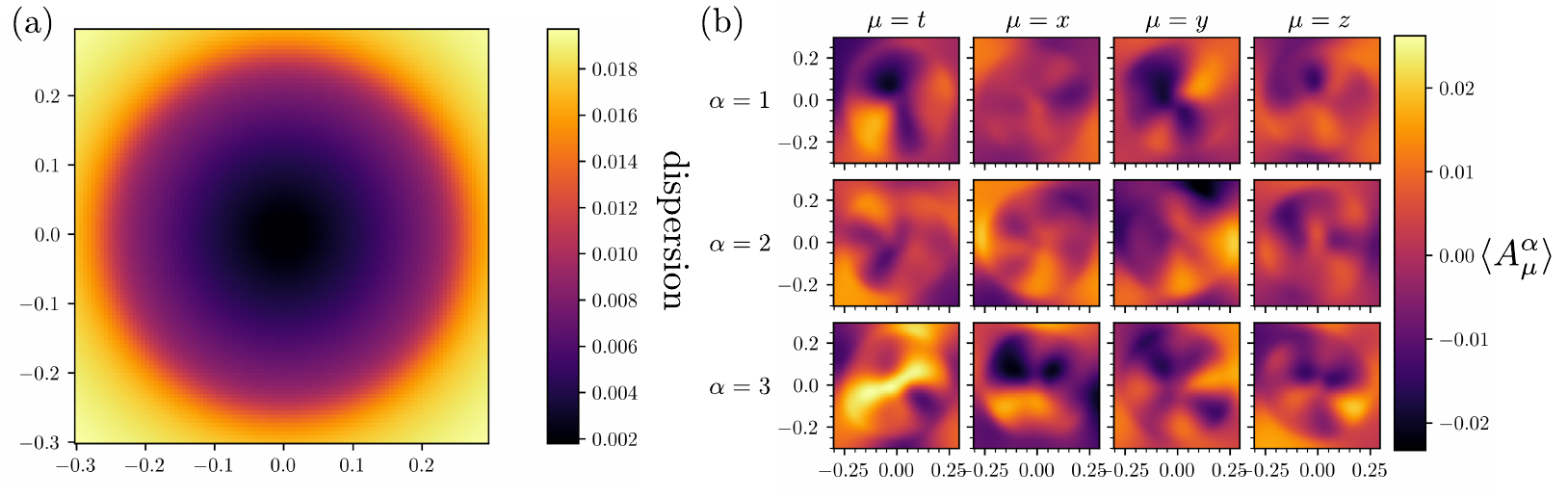}
  \caption{\label{fig:cosm_sim}
    \textbf{$\textit{SU}(2)$ gauge-field observables on an expanding spacetime.}
Results from a Monte-Carlo simulation of a matrix state encoding an $SU(2)$ gauge field on a  $k=-1$ Friedmann-Lema\^itre-Robertson-Walker background. \textbf{(a)} Dispersion map $\Delta^2(\vech{x})$ on a constant-time spatial slice, showing where the matrix state provides a good semiclassical representation of spacetime. The simulation uses eight $500\times500$ Hermitian matrices representing the four spacetime coordinates and their conjugate momenta. Low dispersion is confined to a finite region, reflecting the finite eigenvalue range of the matrices. By varying the time coordinate used to construct the quasi-coherent states $|\vech{x}\rangle$, the evolution of the field on the expanding or contracting background can be analyzed. \textbf{(b)} Corresponding field expectation values $\langle A_\mu^\alpha\rangle = \langle \vech{x} | \text{tr}_n(\tau^\alpha X_\mu) |\vech{x}\rangle$ for the $SU(2)$ gauge field, shown for all spacetime components $\mu$ and field components $\alpha$. The gauge field excitations are introduced by perturbing off-diagonal blocks of the matrix state while keeping the background geometry fixed; the perturbations are sampled element-wise from a Gaussian distribution with variance $\sigma/|\lambda_N|=0.01$. The field is reliably represented only within the low-dispersion region identified in (a), with the diffuse edges reflecting the underlying non-commutative geometry. Together, the panels show that the matrix-model encoding can represent smooth gauge-field dynamics on a curved, time-dependent spacetime while preserving unitary evolution and gauge symmetries.}
\end{figure*}

As a demonstration of the capabilities of matrix model simulations, Fig.~\ref{fig:cosm_sim} presents Monte Carlo results for an $SU(2)$ gauge field on a $k=-1$ FLRW background, encoded in eight $500\times500$ matrices. The dispersion map identifies a finite region of good semiclassical representation, and the gauge-field expectation values $\langle A_\mu^\alpha\rangle$ are smooth within this region, with diffuse boundaries reflecting the underlying non-commutative geometry.

\begin{figure*}
  \includegraphics[width=\textwidth]{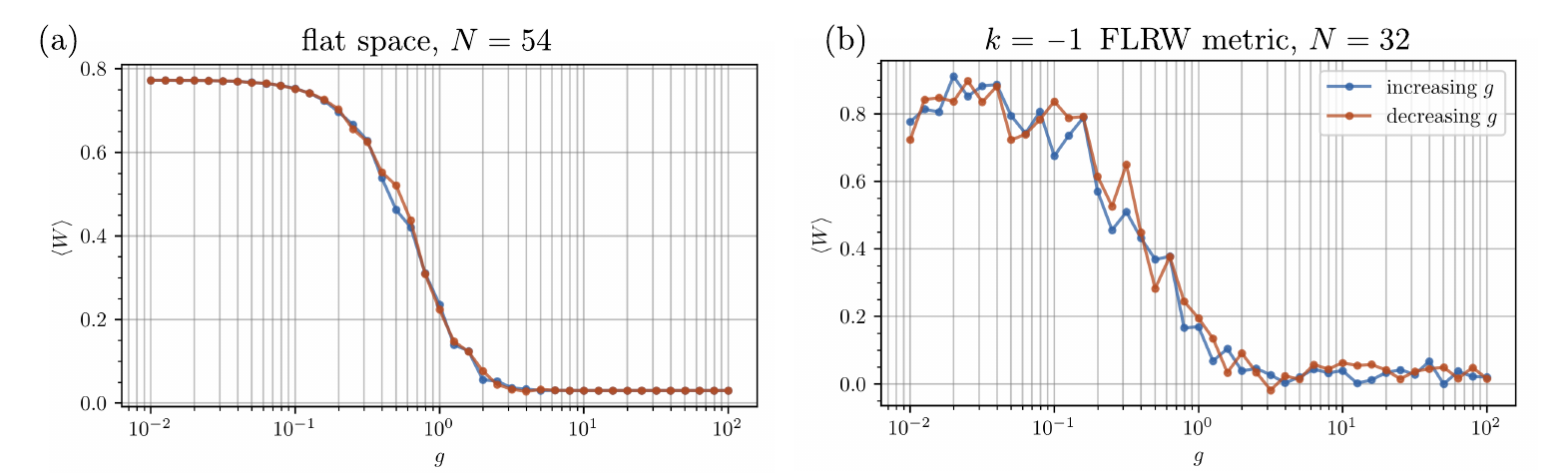}
  \caption{\label{fig:transition-finite}
    \textbf{Deconfinement transition of an $\textit{SU}(2)$ gauge field in flat and curved spacetime.}
Wilson-loop expectation values as a function of the coupling constant $g$ for Monte-Carlo simulations of a Yang-Mills matrix model, for \textbf{(a)} a flat three-dimensional space, with an $\textit{SU}(2)$ gauge field on a $54\times54$ representation of the Moyal-Weyl quantum plane, and \textbf{(b)} for a negatively curved expanding spacetime, with an $\textit{SU}(2)$ gauge field on a $k=-1$ Friedmann-Lema\^itre-Robertson-Walker background using $32\times32$ matrices. The results are evaluated on a single constant-time slice; a smaller matrix size is used in (b) for numerical tractability, though it leads to noisier data. In both panels, the Wilson loop drops from a large value at weak coupling to a much smaller value at strong coupling, indicating a deconfinement phase transition. Forward and backward sweeps of the coupling nearly coincide, showing the minimal hysteresis consistent with a second-order transition.}
\end{figure*}

The central physical demonstration of the framework is presented in Fig.~\ref{fig:transition-finite}, which shows deconfinement phase transitions of $SU(2)$ gauge fields on matrix representations of both flat Euclidean and curved FLRW spacetimes, identified by drops in Wilson-loop expectation values as a function of the coupling constant. 
Wilson loops are evaluated as
\begin{equation}
\langle W_\alpha\rangle=\mathrm{tr}\left(\prod_i
\langle\vech{x}_i|\exp\!\left(i\,\mathrm{tr}_n(\tau_\alpha X_i)\right)
|\vech{x}_i\rangle\right),
\end{equation}
where each $\tau_\alpha\in SU(2)$ is an $N$-dimensional Lie algebra basis element, $X_i$ is the signed matrix for the loop direction at step $i$, and $\mathrm{tr}_n$ is the partial trace over the $SU(n)$-symmetric direct sum structure. The loop is a square of vectors $\vech{x}$ in the $(x_1,x_2)$ plane across a fixed-time Cauchy slice. The Monte-Carlo sampling for the phase transition simulation uses the Metropolis algorithm with a Gaussian proposal distribution. The coupling constant $g$ is swept forward and backward to probe the hysteresis of the $SU(2)$ phase transition.

In both geometries, the Wilson loop falls from a large value at weak coupling to a much smaller value at strong coupling, marking the transition between confined and deconfined phases of the encoded gauge field. Forward and backward Monte Carlo sweeps of the coupling reveal minimal hysteresis, consistent with the second-order transition expected for an $SU(2)$ Yang-Mills theory. 
To our knowledge, this constitutes the first identification of a gauge-field deconfinement transition on an expanding cosmological background within a quantum-simulable encoding, a result inaccessible to lattice gauge theory, which cannot represent the time-dependent metric covariantly.

The fact that these simulations can be evaluated for inflationary spacetimes suggests how these quantum protocols can be applied to cosmological simulations, a category of tasks that is typically difficult to handle with lattice methods \cite{Preskill2012,Liu2021}. These difficulties have even pressed into physical models, 
prompting questions of whether expanding cosmologies can be represented as unitary quantum processes in the first place. \cite{Cotler2022,Brandenberger2022,Brandenberger2001}. However, the existence of matrix encodings that replicate the FLRW metric in the large-$N$ limit suggests a new route to encoding cosmological spacetimes that is consistent with unitary evolution.
Our $SU(2)$ simulation on a $k=-1$ FLRW background in Fig.~\ref{fig:cosm_sim} provides an explicit example of this: unitary evolution is maintained while the spatial slices of the cosmological spacetime are extracted by varying the time parameter used to generate the quasi-coherent states. By taking different Cauchy slices of a covariant matrix state and generating quasi-coherent states across these slices, the field configurations at different points in cosmological expansion can be extracted and compared. 
Field processes and spacetime dynamics can thus be simulated simultaneously within such a matrix model encoding, as the entire spacetime block is represented at once by a single set of matrices $\{X_a\}$ that approach the desired inflationary metrics in the large-$N$ limit. Note that the stability of Lorentzian matrix states within the Euclidean path integral sampled by the protocol involves additional subtleties, for which three mitigation routes are laid out in Appendix \ref{lorentzian-stability-h}. 

These results also clarify the relation between matrix model and lattice encodings. Lattice simulations carry two distinct finite-size errors: discretization errors from the finite lattice spacing, and boundary effects from the finite lattice extent. Matrix models replace the first with the quantum dispersion $\Delta^2(\vech{x})$, spread continuously across the manifold and decaying asymptotically with $N$ (as detailed in Extended Data Fig.~\ref{fig:dispersion_falloff}), and reproduce the second through the finite eigenvalue range of the matrices. The associated error violates neither Lorentz symmetry nor unitarity, in contrast to lattice artifacts. This suggests a natural division of labor: lattice gauge theory remains the method of choice for static, flat-space questions such as hadronic spectroscopy and QCD thermodynamics, while matrix model quantum simulation is most powerful precisely where lattice methods are most strained, for dynamical, curved, or expanding spacetimes and studies of the emergence of geometry from quantum degrees of freedom. A resource estimate for the protocol is provided in Appendix \ref{resource-estimation-g}. The principal requirement is the matrix dimension $N$, since the effective spatial resolution of the encoded geometry is set by the ratio of the total eigenvalue range of the matrix state to the localization radius $\Delta_\text{min}$ of its quasi-coherent states, a ratio that grows with $N$.

Finally, Eq.~(\ref{eq:fid-relation}) generalizes beyond the engineered protocol: any periodic Hamiltonian with phase offsets between matrix components generates commutator terms in its effective Magnus expansion (derived in S.I.~\ref{app:floquet}). Matrix model statistics must therefore arise in a wide class of periodically driven quantum systems without deliberate engineering, as the coherent error of any periodic drive takes on the form of the Yang-Mills action of an effective matrix state. This suggests that observables encoding geometric structures may emerge spontaneously in driven quantum matter, with potential relevance for pre-cosmological models in which structured spacetime emerges from unstructured quantum noise. The framework also connects naturally to holographic approaches to quantum gravity: matrix models such as BFSS and IKKT are conjectured to holographically encode bulk gravitational dynamics, while tensor networks such as MERA \cite{Swingle2012} and holographic codes \cite{Pastawski2015} encode the same dictionary via boundary entanglement structure. By providing quantum-computational access to matrix model path integrals, the present protocol may offer a new route to probing whether the operator-algebraic encoding of bulk geometry can be related to the entanglement-based encoding of tensor networks.

\section{Conclusion}

We have introduced a Floquet framework for quantum simulation of gauge theories that avoids spatial discretization by representing spacetime and field variables entirely through the commutation structure of Hermitian matrix operators. Our central result, that Euclidean matrix path-integral weights are directly accessible as Haar-averaged Floquet fidelities, provides a quantum-native, experimentally plausible route to sampling the partition function of large-$N$ Yang-Mills matrix models. The protocol requires exponentially fewer qubits than canonical quantization, is implementable through standard randomized benchmarking, and naturally accommodates curved and dynamical spacetime geometries including inflationary FLRW backgrounds, on which we have demonstrated the deconfinement transition of an $SU(2)$ gauge field.

Several extensions merit priority. First, fermionic fields can be introduced via the Jordan-Wigner transformation \cite{JordanWigner1928,Verstraete2005} or the natural fermionic statistics of physical platforms, opening the path to supersymmetric matrix models such as the IKKT model \cite{IKKT1997,IKKT1999}, whose one-loop effective action is conjectured to contain an Einstein-Hilbert term \cite{Steinacker2024}. Second, the Lorentzian sector, which we can currently only access via the formal analytic continuation $e^{-S}\to e^{iS}$ during post-processing as detailed in Appendix \ref{lorentzian-stability-h},
can in principle be experimentally encoded on open quantum systems via controlled non-Hermitian operations \cite{Bender1998,Liu2018} equivalent to a Wick rotation. 
Third, matrix states realizing self-intersecting geometries can produce emergent $SU(n)$ gauge fields and multiple generations of chiral fermions \cite{Steinacker2018}, making them attractive targets for quantum simulation of the Standard Model. Deriving matrix states for arbitrary metrics, in particular manifolds of non-uniform curvature such as the Schwarzschild metric, is another crucial avenue for further research.

Altogether, this framework enables entirely new classes of physical simulations. Matrix models encode fields simultaneously with their dynamical background spacetimes, allowing quantum representations of processes, from inflationary expansion to fields near black holes, that have so far been difficult to study outside of classical or semi-classical approximation. Matrix model encodings on quantum systems thus offer not only theoretical insight on how to quantize dynamical spacetime, but also new applications by which experimental quantum platforms can contribute to our understanding of high-energy physics and cosmology. Though the formalism of matrix models is somewhat formidable, its language is deeply compatible with the tools of quantum information, and its constructions offer a wealth of new physics to consider. 

\section*{Acknowledgements}

We are indebted to discussions with Peter Zoller, Ignacio Cirac, Mikhail Lukin, Harold Steinacker, Jordan Cotler, Jacob Barandes, Liyuan Chen, Stefan Ostermann, Hong-Ye Hu, Yidan Wang, and Nik Gjonbalaj. We would like to acknowledge support from the NSF through the CUA PFC (PHY-2317134), the QuSec program (OMA-2326787) and PHY-2207972.

\bibliography{references_nature}

\newpage

\appendix

\section{Encoding fields on matrix geometries}
\label{app:fields-matrix-geometries-a}

To introduce quantum fields on the manifolds represented by sets of matrices $\{ X_a \}$, there are two options, one resulting in scalar field dynamics and the other producing non-commutative Yang-Mills gauge fields.

Starting with scalar fields, we add additional matrices $ \phi_i $ to $\{ X_a \}$, forming the set $ \{ X_a, \phi_i \} $. Plugging this set into the matrix model action results in the expression
\begin{equation}
    S = S[X_a] + \frac{1}{g^2} \text{Tr}\left(2[X^a, \phi^i][X_a, \phi_i] + [\phi^i, \phi^j][ \phi_i, \phi_j]\right).
\end{equation}
When the $X_a$ matrices are fixed and $\phi^i$ are varied, the $[X^a, \phi^i] [X_a, \phi_i]$ kinetic terms act as inner derivations $\eth^a \phi_i \eth_a \phi^i$ on the coadjoint orbits of $X_a$, with $\eth_a \phi_i = \Theta_{ab}^{-1}[X^b, \phi_i]$ as the inner derivation and $\Theta_{ab} = [X^a, X^b]$ acting as the Poisson tensor. As a result, the effective action for fixed $X_a$ adopts the form
\begin{equation}
    S_\text{eff}^\phi = \frac{1}{g^2}\left(  \eth^a \phi_i \eth_a \phi^i + [\phi^i, \phi^j][ \phi_i, \phi_j] \right).
\end{equation}
Translated from the language of matrix models, this is the action of a scalar field on a curved manifold, with quartic interaction terms between all fields.

To introduce Yang-Mills gauge fields possessing $SU(n)$ symmetry, we construct a direct sum of $n$ copies of the matrices on each coordinate, resulting in $U(n)$ invariant matrices $ \bar{T}_a = \bigoplus^n X_a $. When $SU(n)$-symmetric perturbations $T^a = \bar{T}^a + \mathcal{A}^a$ of this direct sum are plugged in to the matrix model Lagrangian, the effective action is that of a non-commutative $SU(n)$ Yang-Mills gauge field. Specifically, the action takes on the form $ S[T_a] = -\frac{1}{g^2} \text{Tr}(\mathcal{F}^{ab} \mathcal{F}_{ab}) $, where the variables in terms of $\Theta^{ab} = [\bar{T}^a, \bar{T}^b]$ are
\begin{equation}
  \begin{cases}
    \mathcal{F}^{ab} = i [T^a, T^b] = -\Theta^{ab} - \Theta^{aa'} \Theta^{bb'} F_{a'b'} \\
    F_{ab} = \eth_a A_b - \eth_b A_a - i [A_a, A_b] \\
    A_a = \Theta^{-1}_{ab} \mathcal{A}^b.
  \end{cases}
\end{equation}
Gauge transformations act on the field variables as
\begin{equation}
  \begin{array}{ll}
    F^{ab} &\rightarrow U^{-1} F^{ab} U \\
    A_a &\rightarrow U^{-1} A_a U + U^{-1} i \eth_a U.
  \end{array}
\end{equation}
From this unitary symmetry, $F^{ab}$ acts as the field strength tensor of an $SU(n)$ non-commutative Yang-Mills field. Under fixed $\bar{T}^a$, the effective action is the Yang-Mills field action $S_\text{eff}^\text{YM} = -\frac{1}{g^2} \text{Tr}\,F_{ab} F^{ab}$, with indices contracted by the curved metric $g_{ab}$ induced by $\bar{T}^a$. Gauge potential expectation values are extracted in post-processing as
\begin{equation}
  \langle A_a^{\alpha}(\vech{x}) \rangle = \langle\vech{x}|\, \mathrm{tr}_n(\tau_\alpha T_a)\, |\vech{x}\rangle,
\end{equation}
where $\tau_\alpha$ is a basis element of the $SU(n)$ Lie algebra, $T_a$ is the element of the matrix state for the desired spacetime direction, and $\mathrm{tr}_n$ is the partial trace over the $SU(n)$ direct-sum structure used in constructing $\bar{T}_a$. A more comprehensive treatment of these field encodings, including the matrix energy-momentum tensor, the Schwinger-Dyson equation serving as the Ward identity that replaces Gauss's law constraints, and the coupling of scalar and gauge sectors, is provided in Supplementary Information (S.I.)~\ref{app:fields}.

\section{Quasi-coherent state post-processing}
\label{quasicoherent-b}

Each result generated with a given quasi-coherent state $|\vech{x}\rangle$ of Eq.~(\ref{eq:quasi-hamiltonian}) possesses a finite uncertainty, or dispersion, associated to each classical position $\vech{x}$,
\begin{equation}
  \Delta^2(\vech{x}) = \sum_{a=1}^d \langle \vech{x} | X^a X^a | \vech{x} \rangle - \langle \vech{x} | X^a | \vech{x} \rangle^2.
\end{equation}
Another function on the space of quasi-coherent states is the displacement, measuring the error between the expectation value of the quasi-coherent state and its intended classical value,
\begin{equation}
  d^2(\vech{x}) = \sum_{a=1}^d (\langle \vech{x} | X^a | \vech{x} \rangle - x^a )^2.
\end{equation}
Both of these measures can be used to study the emergence of classical geometries from quantized Hermitian operators, and both are directly related to the quasi-coherent state eigenvalue $H_{\vech{x}} |\vech{x}\rangle = \lambda_0(\vech{x})|\vech{x}\rangle$ in terms of the identity $2 \lambda_0(\vech{x}) = \Delta^2(\vech{x}) + d^2(\vech{x})$.

For a typical set of matrix variables, there will be a subset of classical vectors $\vech{x}$ whose dispersions and displacements are minimal. This subset represents the classical spacetime manifold encoded by the matrix variables, whose dimension may be less than the number of matrices. For instance, the non-commutative sphere $\{r J^{(N)}_a \}$ is a two-dimensional manifold defined in terms of three matrices, representing three embedding dimensions for the classical manifold $S^2$. The induced metric on such embedded manifolds is given by
\begin{equation}
g^{ab}(\vech{x})=\langle\vech{x}|\delta_{a'b'}[X^a,X^{a'}][X^b,X^{b'}]|\vech{x}\rangle,
\end{equation}
derived in S.I.~\ref{app:geometries}. We provide visual plots of $\Delta^2(\vech{x})$ and $d^2(\vech{x})$ alongside their associated classical manifolds for a variety of illustrative matrix states in Fig.~\ref{fig:hexaplot}, and elaborate on their definition and representation as finite matrices in S.I.~\ref{app:jordan-schwinger}.

\section{Floquet sequence compilation and validity}
\label{floquet-derivation-c}

We extended the Floquet sequence across an entire set of $d$ matrices $\{X_a\}$ in Eq.~(\ref{eq:fid-relation}) by applying sequences for every pair of operators by hand and then Trotterizing these together, for a total of $4(d^2-d)/2$ operations. An efficient alternative to this naive Trotterization, requiring only $\mathcal{O}(d)$ sequential operations is to generate the same sum of commutators with sequential forward and backward sequences of unitary operations, such that
\begin{align*}
  \mathcal{F}_2(\{X_a\}; \Delta t) &= \left( \prod_a e^{i \Delta t X_a} \right) \left( \prod_b e^{-i \Delta t X_b} \right) \numberthis \\
  &= \exp\left( - i \Delta t^2 \sum_{a,b} i [X_a, X_b]  + \mathcal{O}(\Delta t^3)\right),
\end{align*}
where care must be taken to always apply the forward and backward sequences of operators in the same order. 
The full BCH expansion of the sequence, the trace expansion underlying Eq.~(\ref{eq:frob-haar}), and the higher-order error terms are derived in S.I.~\ref{app:floquet}.

For the leading-order identification of fidelities with path-integral weights to hold, the higher-order terms must vanish as $\Delta t^6\,\text{Tr}([X_a, [X_a, X_b]]^2) \ll 1$. Note that for studying spacetime configurations, matrix models are best put in the weak coupling regime where $g \ll 1$ and classical configurations dominate, while non-local higher-spin gauge field dynamics prevail in the strong coupling regime \cite{Steinacker2024}.

Evaluating $\langle f\rangle$ via Eq.~(\ref{eq:weingarten}) is equivalent to global $U(N)$ twirling of $\mathcal{F}_2$ followed by gate fidelity estimation \cite{Emerson2005,Knill2008,Magesan2011}, treating the commutator terms $[X_a,X_b]$ as the only source of coherent error. Approximate Haar sampling via Clifford 2-designs is sufficient for evaluating this operator fidelity \cite{Dankert2009}, making the protocol directly compatible with established randomized benchmarking methods on digital quantum platforms, without needing to sample directly from the full qudit Haar measure.

The same statistics arise from generalized periodic driving. Any periodic Hamiltonian alternating between two operators, such as $H(t) = \sin(\omega t) X_a + \cos(\omega t) X_b$, has a vanishing first-order Magnus term and a second-order term $\Omega_2 = -\frac{\pi}{\omega^2}[X_a, X_b]$, yielding the effective Hamiltonian $H_\text{eff} = -\frac{1}{2\omega}[X_a, X_b]$ in the high-frequency limit. Floquet driving for these simulations therefore does not need to be engineered with exact sequences of successive operators, but can arise from any periodic driving with amplitude modulation at offset phases distributed across the unit circle. The corresponding derivation for arbitrary matrix states is given in S.I.~\ref{app:floquet}.

\section{Globally-controlled parallel circuit}
\label{globally-controlled-d}

Operator product expansions for the matrix model, defined by Eq.~(\ref{eq:ope}), are evaluated in the quantum simulation via controlled gates across an ensemble of Floquet sequences, with the architecture illustrated in Fig.~\ref{fig:parallel-circuit}a. Each Floquet sequence $\mathcal{F}_2(\{X_a\}_k; \Delta t)$ is conditioned on the computational basis state $|B_k\rangle$ of the control qubits via a multi-qubit controlled unitary gate, which in a gate-based implementation reduces to a network of Toffoli (CCU) gates.

The basis of $n$ control qubits is composed of elements $|B_k\rangle$ notated as
\begin{equation}
  |B_k\rangle = \bigotimes_{i = 1}^n |b_{i,k}\rangle,
\end{equation}
where $b_{i,k}$ is the $i$th digit of the binary representation of the number $k$. In the parallelized circuit, the basis $\{ |B_k\rangle \}$ acts as a binary index for the matrix model states $\{X_a\}_k$ and their corresponding operators $\mathcal{O}[\{X_a\}_k] = \mathcal{O}_k$ in the operator product expansion under consideration, over $2^n$ possible values of $k$. Controlled gates are set up such that the $k$th control qubit state activates the Floquet sequence $\mathcal{F}_2(\{X_a\}_k; \Delta t)$ applied to a Haar-random target state $|\psi\rangle$.

After the Floquet sequences controlled by the entangled state $|\mathcal{B}\rangle$ are applied to the Haar-random state $|\psi\rangle$, the state transforms on average from $|\psi\rangle |\mathcal{B}\rangle$ to
\begin{align}
  \left( \frac{1}{\sqrt{2^n}} \sum_{k} \sum_{(X_a^{(k)}, X_b^{(k)})} e^{i \theta_k} e^{-\frac{1}{2} \text{tr}([X_a^{(k)}, X_b^{(k)}]^2)} |\psi\rangle |B_k\rangle\right) \\ 
  + \left( \frac{1}{\sqrt{2^n}} \sum_k \alpha_k |\psi_\perp\rangle |B_k\rangle \right), \numberthis 
\end{align}
where $(X_a^{(k)},X_b^{(k)}) \subset \{X_a\}_k$ denotes the pairs of operators drawn from the matrix states $\{X_a\}_k$ whose Floquet sequence is activated by control qubit state $|B_k\rangle$, and the target state $|\psi_\perp\rangle$ denotes all states orthogonal to the original Haar-random state $|\psi\rangle$.

Under each Floquet sequence, the fidelity for the final target state remaining in its initial state $|\psi\rangle$ after the measurement is
\begin{equation}
  \langle f_k \rangle = \int d\mu(\psi) |\langle \psi | \mathcal{F}_2(\{X_a\}_k; \Delta t) |\psi\rangle|^2 = e^{-S[\{X_a\}_k]}.
\end{equation}
Post-selecting on the measurement outcomes in which the target states remain at their initial states $|\psi\rangle$ traces out the $|\psi_\perp\rangle$ branch, leaving the final control qubit state as
\begin{equation}
  \frac{1}{\sqrt{2^n}} \sum_k \sum_{(X_a^{(k)}, X_b^{(k)})} e^{i \phi_k} e^{-\frac{1}{2}\text{tr}([X_a^{(k)}, X_b^{(k)}]^2)} |B_k\rangle.
\end{equation}
Reading off the corresponding control qubit outcomes then provides an ensemble of relative frequencies $\langle f_k \rangle$ across the control qubit indices $|B_k\rangle$, weighted relative to each other by Eq.~(\ref{eq:pk}) and thus reproducing the probability distribution of the Euclidean path integral. An appropriately weighted operator product expansion is calculated from these probabilities as $\langle \mathcal{O} \rangle = \sum_k p_k \mathcal{O}_k$, evaluating any operator product expansion of the matrix model path integral across a fixed discretization of the measure indexed by $k$, in parallel across the even superposition $|\mathcal{B}\rangle$.

Experimentally, Haar-random states are generated by starting with a known state $|0\rangle$ and applying a fixed random unitary $U$ from the Haar measure. The post-selection of the state after Floquet sequences are applied can then be simply achieved by applying the inverse $U^\dagger$, and measuring whether the target qubits are in the state $|0\rangle$.

\section{Locally-controlled circuit}
\label{locally-controlled-e}

The globally controlled circuit offers a general protocol for sampling from a matrix model path integral, but it requires high-order Toffoli gates across arbitrarily entangled global unitary operations. Such general features are not only difficult to implement on experimental quantum platforms, but are also not necessary to achieving an effective matrix model quantum simulation with Floquet driving. As an alternative, we construct a locally-compiled parallel quantum circuit, with perturbing sparse unitary operations interspersed in a global Floquet drive. Each operation is controlled by a single qubit rather than the entire multi-qubit basis, avoiding the need for coupling to delicate entangled states.

\begin{figure*}
  \includegraphics[width=\textwidth]{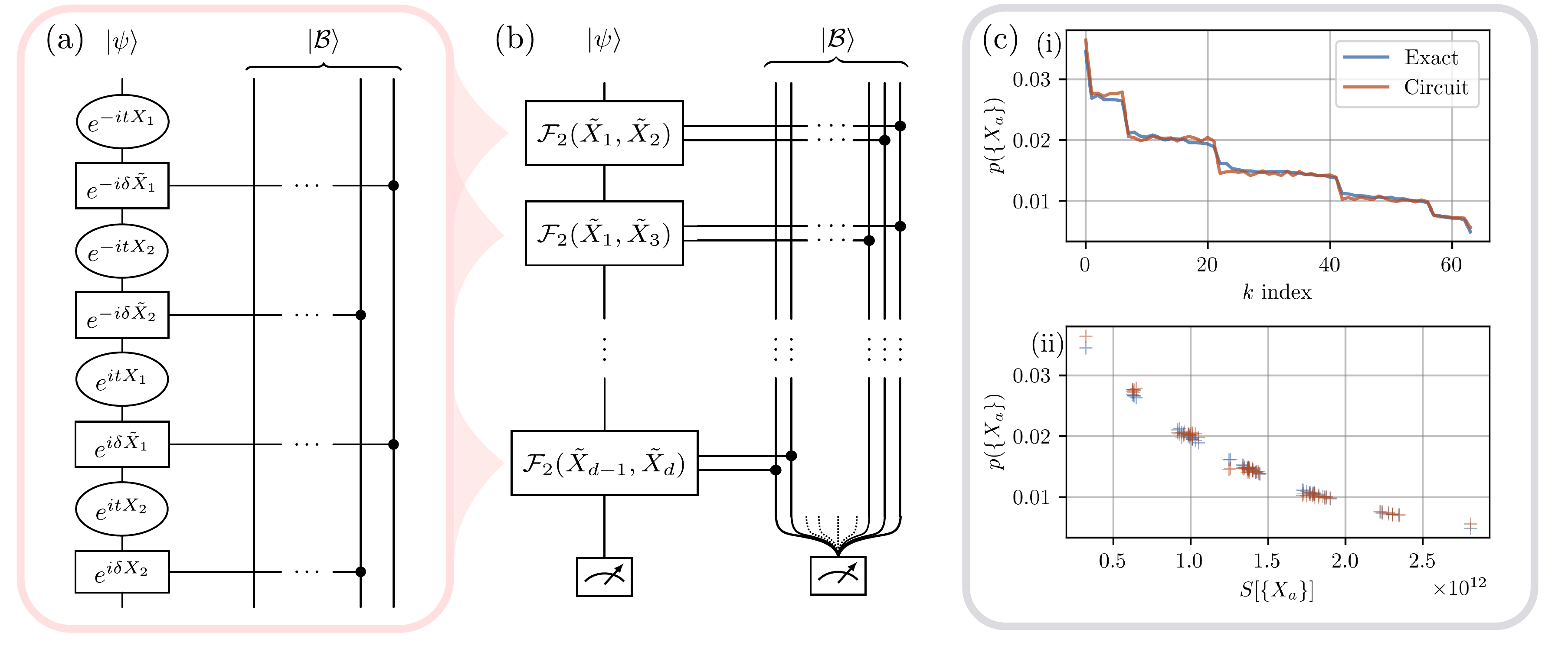}
  \caption{\label{fig:scrambling-circuit}
     \textbf{Locally controlled quantum circuit and path-integral comparison (Extended Data).}
\textbf{(a)} Diagram of the locally controlled parallel circuit described in the Appendix \ref{locally-controlled-e}. For a bare Floquet sequence $\mathcal{F}_2(\{X_a\}; \Delta t )$ generated from a matrix state $\{X_a\}$, perturbations $\delta\tilde{X}_a$ to the individual matrix elements are controlled by separate control qubits. For each four-step sequence $\mathcal{F}_2(X_a,X_b; \Delta t )$, the corresponding control qubit inserts unitary perturbations of the form $e^{\pm i\delta \tilde{X}_a}$ into the sequence. Simultaneous perturbations of multiple elements arise from the entangled branches of the control state $|\mathcal{B}\rangle$.
\textbf{(b)} The full circuit across all Floquet sequences for elements of the matrix state $\{X_a\}$. In contrast to Fig.~\ref{fig:parallel-circuit}, control over sampling the simulated path-integral measure is localized to small sets of control qubits, rather than requiring Toffoli gates across the full entangled basis.
\textbf{(c)} Comparison of fidelities from a numerical simulation of the locally controlled circuit with an ideal Monte-Carlo simulation of the Euclidean path integral over the same ensemble. As in Fig.~\ref{fig:parallel-circuit}, the matrix states are perturbations of the Moyal-Weyl quantum plane, here represented by $10\times10$ matrices for numerical stability. Element-wise Gaussian perturbations are drawn up to a maximal variance $\sigma/|\lambda_N|=0.1$. Floquet cycles are applied at $\hbar\omega_F/|\lambda_N|=\hbar/(t|\lambda_N|)=\hbar/(\delta|\lambda_N|)=4500$ over 325000 cycles. Three perturbations are included for each of the four matrix elements, for a total of 12 control qubits. After post-selection on stationary target-qubit outcomes, the probabilities of the control-basis states are shown in (i) as a function of the basis index $k$, together with the exact Euclidean weights obtained from the corresponding perturbed matrix states. Because the locally controlled circuit is noisier than the globally controlled version, the results are averaged over 100 trials. The plateau structure reflects the discrete perturbations introduced as additional control qubits are activated. In (ii), the same data are plotted against the action $S[\{X_a\}]$, showing the expected exponential weighting of the Euclidean matrix path integral.}
\end{figure*}

In the setup presented in Extended Data Fig.~\ref{fig:scrambling-circuit}, each individual control qubit $|b_i\rangle$ activates distinct operations $U_b = e^{- i \delta \tilde{X}_a}$ for a perturbing matrix $\tilde{X}_a$, applied periodically and in phase with each initial matrix operator ${X}_a \in \{ {X}_a \}$, so that the components of the Floquet sequence involving ${X}_a$ and any other matrix ${X}_b$ are modified from $ \mathcal{F}_2({X}_a, {X}_b; \Delta t) =  e^{i \Delta t X_b} e^{i \Delta t X_a} e^{-i \Delta t X_b} e^{-i \Delta t X_a} $ to
\begin{equation}
  e^{i \Delta t X_b} e^{i \delta \tilde{X}_a} e^{i \Delta t X_a}   e^{-i \Delta t X_b} e^{-i \delta \tilde{X}_a} e^{-i \Delta t X_a}.
\end{equation}
With this circuit arrangement, each control qubit only has to activate a single fixed operator $e^{\pm i \delta \tilde{X}_a}$, rather than the more challenging setup in which entangled states control distinct global Floquet operations.

While the time interval $ \delta $ may be larger than $\Delta t$, their sum must satisfy $ \Delta t + \delta \ll 2\pi / \|\tilde{X}_a\| $ to preserve the validity of Eq.~(\ref{eq:fid-relation}). Only the ratio $ \delta/\Delta t $ should be varied, while keeping their total periods small. For $\delta \ll t$, this step of the sequence approaches $ \mathcal{F}_2(X_a + \frac{\delta}{\Delta t} \tilde{X}_a, X_b; \Delta t) $, and the circuit samples from a narrow radius off-shell around the matrix solution generated by $\mathcal{F}_2(\{X_a\}; \Delta t)$. Alternatively, for driving over time intervals $ \delta \gg \Delta t $, the higher order terms in the Baker-Campbell-Hausdorff expansion of $e^{-i \delta \tilde{X}_a} e^{-i \Delta t X_a}$ become relevant to the dynamics, and the matrix state $\{X_a\}$ is scrambled in a nonlinear fashion. The full operator space is covered by the locally-controlled operations, enabling uniform sampling over the whole matrix model path integral measure. In particular, when multiple operations are activated at once by control qubits in the large-$\delta$ regime, nonlinear interactions between the locally-controlled operators arise in the effective Hamiltonian via higher-order BCH terms, taking the effective matrix state arbitrarily far outside the coadjoint orbit of the initial matrix state $\{ X_a\}$. This setup exhibits a tradeoff between computation time and depth of sampling across the measure.

\section{Numerical parameters}

For Fig.~\ref{fig:fidelities}, the matrix states used in the numerical simulations of the Floquet sequence $\mathcal{F}_2$ are $100\times100$ Hermitian matrices encoding the Moyal-Weyl quantum plane, perturbed element-wise by Gaussian noise across 50 evenly-spaced increments of variance from zero to $\sigma/|\lambda_N|=1.5$, where $|\lambda_N|$ is the maximal eigenvalue. At each increment, 100 perturbed matrix states are sampled. For each state, the Floquet sequence $\mathcal{F}_2$ is applied to 100 Haar-random states, yielding $\langle f\rangle$. The results at three Floquet frequencies are shown ($\hbar\omega_F/|\lambda_N|=3.3\times10^4$, $1.0\times10^5$, $3.3\times10^6$). Residual deviation at high frequency is dominated by the $\mathcal{O}(\Delta t^3)$ BCH remainder.

The $k=-1$ FLRW simulation presented in Fig.~\ref{fig:cosm_sim} uses eight $500\times500$ Hermitian matrices representing four spacetime coordinates and their conjugate momenta, constructed from generators of $SO(4,2)$ as described in S.I.~\ref{app:jordan-schwinger}. The gauge field is introduced by perturbing off-diagonal blocks with element-wise Gaussian noise up to a variation of $\sigma/|\lambda_N|=0.01$. Quasi-coherent states are constructed for a grid of spatial positions at a fixed time coordinate to provide visuals for the gauge potential expectation values across the target space.

For Fig.~\ref{fig:parallel-circuit}, the control qubits encode 32 matrix states from the $100\times100$ Moyal-Weyl plane, each perturbed by element-wise Gaussian noise at a variety of variances from zero up to $\sigma/|\lambda_N|=0.02$. The driving frequency is $\hbar\omega_F/|\lambda_N|=224$ over 1375 Floquet cycles. For Extended Data Fig.~\ref{fig:scrambling-circuit}, the locally-controlled circuit uses $10\times10$ matrices for numerical stability. The circuit makes use of 12 control qubits, activating three distinct perturbations over four matrix elements. The driving frequency is $\hbar\omega_F/|\lambda_N|=4500$ over 325{,}000 cycles, and results are averaged over 100 trials.

Finite-size effects in finite-dimensional matrix states are quantified in Extended Data Fig.~\ref{fig:dispersion_falloff} by the minimum displacement $d^2(\vech{x})$ and dispersion $\Delta^2(\vech{x})$ of quasi-coherent states for spherical matrix geometries with $N=10$ to $200$, with the matrices normalized to fixed Frobenius norm $\|X_a\|^2=\mathrm{Tr}(X_a^2)=1$ so that results can be compared across dimensions. Both the encoding error and the uncertainty in spatial information decrease asymptotically with increasing matrix dimension.

\section{Resource estimation}
\label{resource-estimation-g}

\begin{figure}
  \includegraphics[width=\linewidth]{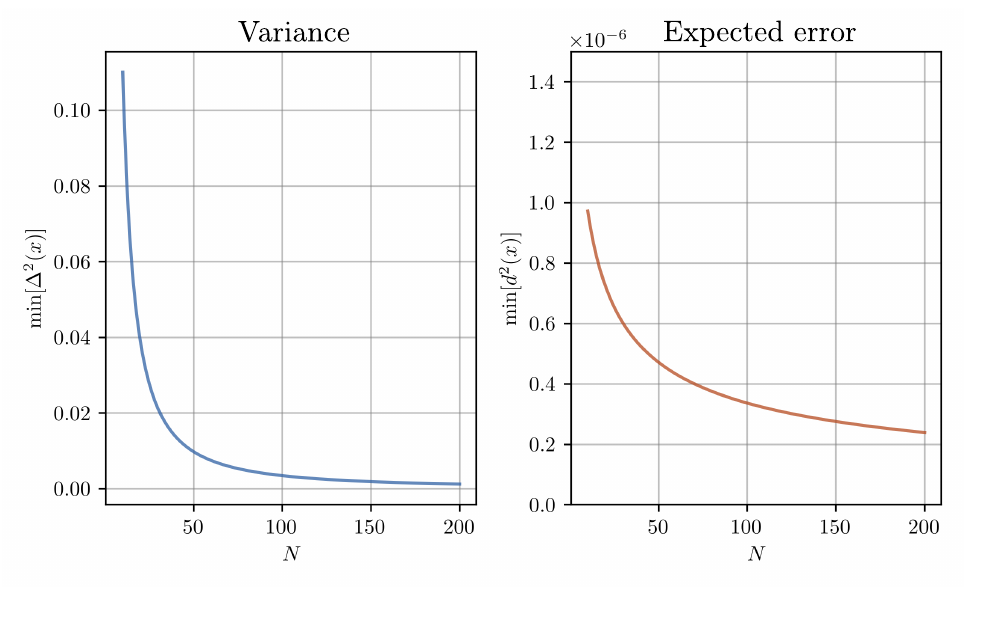}
  \caption{\label{fig:dispersion_falloff}
    \textbf{Matrix scaling and finite-size effects (Extended Data).}
Finite-size effects in finite-dimensional matrix states, quantified by the minimum displacement $d^2(\vech{x})$ and dispersion $\Delta^2(\vech{x})$ of quasi-coherent states. The matrix states encode a spherical geometry using $N\times N$ matrices with $N=10$ to $200$. For each $N$, the matrices are normalized to fixed Frobenius norm, $||X_a||^2=\mathrm{Tr}(X_a^2)=1$, so that results can be compared across dimensions. Both the encoding error and the uncertainty in spatial information decrease asymptotically with increasing matrix dimension.}
\end{figure}

The cost of the Floquet protocol is controlled by three parameters. The first is the Floquet timestep $\Delta t$, which must satisfy $\Delta t|\lambda_N| \ll 1$ for the leading-order identification of fidelities with path-integral weights to remain valid, where $|\lambda_N|$ is the maximal eigenvalue of the matrix state. Through the identification $1/g^2 = 2\Delta t^4/(N+1)$, the timestep also fixes the effective coupling of the simulated matrix model. The second parameter is the total drive time, equal to the number of Floquet cycles times the timestep. Because the sampled weight of a configuration falls off as $e^{-S[X]}$ and a configuration is resolvable only while its weight remains above the Haar overlap floor of order $1/N$, the total drive time sets how finely actions can be distinguished, and together with the timestep determines the range of effective couplings that a given run can probe. The third parameter is the matrix dimension $N$, which determines the spatial resolution of the encoded geometry. This resolution is the number of quasi-coherent states distinguishable across the manifold, given by the ratio of the total eigenvalue range of the matrix state to the localization radius 
$ r_\text{loc} = \sqrt{\text{min}[\Delta^2(x)]} $ across the available quasi-coherent states for a given matrix configuration $\{X_a\}$. 
Since $r_\text{loc}$ decreases asymptotically with $N$, as shown in Extended Data Fig.~\ref{fig:dispersion_falloff}, finer geometries are reached by increasing the matrix dimension, at a cost set by the encoding of larger Hamiltonians.

The circuit depth needed to implement each Floquet unitary is typically modest, as the background matrix operators are sparse. The $SU(2)$ angular momentum generators, the finite ladder operators of the Moyal-Weyl plane, and the $SO(4,2)$ generators of the FLRW background are all banded, with a constant number of non-zero entries per row, so each $e^{\pm i \Delta t X_a}$ admits an efficient gate decomposition by standard sparse Hamiltonian simulation methods~\cite{Barenco1995}. The perturbations $\delta\tilde{X}_a$ of the locally controlled circuit are generally cheaper still. Random element-wise perturbations require no structured compilation, and the low-energy eigenmodes of the background matrix state furnish a sparse parametrization of the physically relevant off-shell configurations. As a concrete worst case, the $k=-1$ FLRW simulation reported here uses eight $500\times500$ matrices, so each Floquet unitary acts on $2\log_2 500 \approx 18$ qubits, with the number of cycles fixed by the ratio $|\lambda_N| / r_\text{loc}$ ratio at this dimension and the per-cycle depth set by the sparsity of the eight operators.

\section{Stability of Lorentzian matrix states in the Euclidean path integral}
\label{lorentzian-stability-h}

The Floquet fidelity protocol naturally produces a Euclidean matrix model action $S_E = -\frac{1}{g^2}\text{tr}(\delta_{aa'}\delta_{bb'}[X^a,X^b][X^{a'},X^{b'}])$, while Lorentzian matrix states are stationary points of the Minkowski-signature action $S_L$ in which $\delta_{aa'} \delta_{bb'}$ is replaced by $\eta_{aa'} \eta_{bb'}$. Sign discrepancies therefore arise for commutators of the form $[X^0, X^i]$, and Lorentzian saddle points are not in general saddle points of $S_E$. Three independent routes are available to address this. First, the matrix state can be Wick-rotated, replacing $SO(4,2)$ generators with those of $SO(6)$ via $X^0 \to e^{i\pi/2}X^0$ \cite{Nishimura2022}; observables are then analytically continued back to Lorentzian signature during post-processing. Second, Lorentzian geometric fluctuations can be restricted to the perturbative saddle point of the Euclidean action, corresponding to the $\delta \ll t$ limit of the locally-controlled circuit described above, where non-geometric excitations are suppressed and an effective metric $g_{ab}$ remains well-defined throughout. Third, the part of the matrix state encoding the background spacetime can simply be held fixed, so that the path integral runs only over field configurations; in this case the $[X^0, X^i]$ commutators are constants rather than integration variables, and their sign difference between $S_L$ and $S_E$ contributes a fixed overall factor to the measure rather than a configuration-dependent instability.

\section{Experimental considerations}

The quantum circuits described above are hybrid quantum simulation protocols, consisting of an analog component, the Floquet sequences $\mathcal{F}_2$ for each matrix state, and a digital component, the control qubits $|\mathcal{B}\rangle$ for the simulation readout. Regardless of which exact circuit setup one uses, the quantum operations for any quantum simulation of matrix models in terms of fidelity measurements will consist of high-dimensional unitary operations controlled and reversed by individual control qubits and acting on sets of target qubits, followed by the measurement of the Loschmidt echo of those target qubits. This setup can be achieved particularly naturally by an experimental configuration devised for quantum non-demolition measurement of many-body Hamiltonians \cite{Zoller2020_NatureComm,Zoller2020_PRXQ}, in which the target qubits are placed within the Rydberg radius of a handful of control qubits, as either a 1D circle or a 2D lattice. The control qubit state $|\mathcal{B}\rangle$ and Haar-random target qubit state $|\psi\rangle$ can be generated in separate physical locations within the experiment, and the atoms can then be arranged into a computational configuration within their respective Rydberg radii. The differentiation of the control qubit and target qubit interaction regimes in such a geometry can be particularly favorable on dual-species atomic arrays \cite{Bernien2024, Semeghini2025}. The available Hilbert space dimensions for the simulated matrices are then dictated by the number of target qubits that can be packed into the Rydberg radius of the control qubits. By placing the control qubits in GHZ states and using these collective states as the control basis, the region of Rydberg excitation can be made to cover arbitrarily many target qubits. We offer this description not as a complete account, but as a sketch of a viable experimental setup for the quantum circuits considered here.

\setcounter{section}{0}

\newpage 
$ $ 
\newpage 
\onecolumngrid

\section*{Supplementary Information}

Supplementary Sections \ref{app:quasicoherent}--\ref{app:haarrelations}, included as appendices in this manuscript file, contain the quasi-coherent state formalism, the derivation of effective geometries from operator algebras, the finite Jordan-Schwinger constructions of the matrix geometries and field encodings, the full Floquet sequence and generalized periodic driving derivations, and the Weingarten calculus proof of the Haar-fidelity identity.

\subsection{Quasi-coherent states and dequantization}
\label{app:quasicoherent}

Matrix states consisting of Hermitian operators $\{X_a\}$ offer a completely global description of a spacetime and its fields; we would like a consistent procedure for extracting local information about the classical manifolds that they encode. 

We can start with the most straightforward case, where $\{X^a\} = \{Q^i, P^i\}$ for $d$ canonical position and momentum operators. 
The procedure for mapping classical points in the phase space to a quantum approximation with minimal uncertainty is straightforward. 
We use coherent states $| \vech{x}\rangle$, where $\vech{x}$ is a vector of position and momentum coordinates. The state $|\vech{x}\rangle$ is defined as the state with minimal uncertainty between operators $Q^i$ and $P^i$ at point $\vech{x}$. 
Equivalently, $|\vech{x}\rangle$ can be defined as the ground state of a simple harmonic oscillator whose resting point in the phase space has been displaced by the vector $\vech{x}$ consisting of real-valued entries $x^a$. 
Coherent states $| {x}\rangle$ are then the ground states of the following Hamiltonian: 
\begin{equation} \label{quasi-Hamiltonian}
  H_{\vech{x}} = \frac{1}{2} \sum_{a=1}^d \left( X^a - x^a \openone \right)^2.
\end{equation}
From this perspective, we can easily generalize the formalism for coherent states to any set of operators $\{ X^a \}$. We refer to such generalized coherent states for arbitrary matrix operators ${X^a}$ as \textit{quasi-coherent states}. These quasi-coherent states are normalized to 1 and live in the eigenspace of $H_{\vech{x}}$ with eigenvalue 
$H_{\vech{x}} |\vech{x}\rangle = \lambda_0(\vech{x}) |\vech{x}\rangle$. 

Using these quasi-coherent states, we define a de-quantization map that maps from noncommutative observables to classical functions, 
\begin{equation} 
  \mathcal{O}\to \langle \vech{x}|\mathcal{O}|\vech{x}\rangle = \mathcal{O}(\vech{x}).
\end{equation}
The expectation values $\langle \vech{x} | \hat{X}_a |\vech{x}\rangle = \tilde{x}_a$ define a new vector $\tilde{\vech{x}}$, representing the best approximation of the original $\vech{x}$ that the operators $\{X_a\}$ can produce. 
In general, this new vector is not identical to $\vech{x}$, and the difference between them is an important measure of how classical 
the non-commutative manifold encoded by $\{ X_a \}$ really is. The distance between $\vech{x}$ and $\tilde{\vech{x}}$ is termed the \textit{displacement}, and can be evaluated as 
\begin{equation}
  d^2(\vech{x}) = \sum_{a=1}^d (\langle \vech{x} | X^a | \vech{x} \rangle - x^a )^2.
\end{equation}
A second important measure on the space of quasi-coherent states is the uncertainty inherent at each point on the manifold. This is given by the \textit{dispersion}, which mimics the formula for Heisenberg uncertainty at a point, 
\begin{equation}
  \Delta^2(\vech{x}) = \sum_{a=1}^d \langle \vech{x} | X^a X^a | \vech{x} \rangle - (\langle \vech{x} | X^a | \vech{x} \rangle)^2.
\end{equation}
Both of these measures indicate the exact vectors $\vech{x}$ for which there is an emergence of classical geometries from the overall matrix state. The relation $2 \lambda_0(\vech{x}) = \Delta^2(\vech{x}) + d^2(\vech{x})$ makes clear that minimizing the ground state energy of $H_{\vech{x}}$ also minimizes $\Delta^2(\vech{x})$ and $d^2(\vech{x})$, and thus picks out the quantum states $|\vech{x}\rangle$ most closely associated to classical phase space coordinates $\vech{x}$ on some underlying manifold. 

These measures of classicality are typically not small for all $\vech{x}$. The submanifold of $\vech{x}$ vectors for which these measures are small indicates the manifold that the matrix state $\{X_a\}$ approaches in the large-$N$ limit. In particular, all low-energy excitations of the state $\{X_a\}$ within a Yang-Mills matrix model can be decomposed into harmonics that propagate on this optimally classical submanifold \cite{Aoki1998,Iso2000,Steinacker2010,Steinacker2024}. 
This is the mechanism through which matrix models encode field excitations on arbitrary curved classical spacetime manifolds, as well as any desired geometric fluctuations of those manifolds themselves.

\subsection{Operator algebras and effective geometries}
\label{app:geometries}

Returning to the flat geometry of canonical position and momentum operators as our fundamental case, we now discuss how to extract Poisson structures and metric tensors from quantum operator algebras. 
The matrix Laplacian indicates that there is a flat Euclidean geometry associated with any set of $2n$ canonical position and momentum operators $X^a \in \{ Q^i, P^i \}$ obeying the commutation relations, $[Q^i, P^j] = i \hbar \delta^{ij} \openone$. This commutation relation can be rewritten in terms of a fully anti-symmetric tensor,
\begin{equation}
  \Theta^{ab} = \begin{pmatrix}
    0 & -\openone \\ 
    \openone & 0
  \end{pmatrix}
\end{equation} 
which encodes the Poisson structure of the position and momentum operators. The commutator in terms of $ X^a $ is then
\begin{equation}
  [X^a, X^b] = i \Theta^{ab} \openone.
\end{equation}
As a result, the matrix Laplacian applied to any of its generators always returns zero, 
\begin{equation}
    \square X^a = [X_a, [X^a, X^b]] = [X^a, i \hbar \delta^{ab} \openone ] = 0.
\end{equation}
If we interpret the matrix Laplacian as measuring a curvature of an underlying manifold, then this result implies that the manifold is flat. The geometry associated with these canonical operators is referred to as the Moyal-Weyl quantum plane \cite{Moyal1949,Weyl1931,Groenewold1947}.

This interpretation is motivated by the fact that
a single commutator behaves analogously to a derivative. In particular, a differentiable function $\Phi(X^a)$ of operators $X_a$ can be written as a power series,  
\begin{equation}
  \Phi(X^a) = \sum_a \sum_{k=0}^\infty \frac{\alpha_{a, k}}{k!} (X^a)^k.
\end{equation}
By defining a canonical derivative $\eth_a$ in terms of a commutator as, 
\begin{equation} \label{canonical_deriv}
  \eth_a \Phi = -i \Theta_{ab}^{-1} [X^b, \Phi] 
\end{equation}
we can extract the coefficients $\alpha_{a,j}$ by nesting commutators, just as if we were taking derivatives on a Maclaurin series: 
\begin{equation}
  (\eth_a)^k \bigl( \Phi(X^a) \bigr) = \alpha_{a, k} \openone + \mathcal{O}(X^a).
\end{equation}
Strictly speaking, $\eth_a$ is a derivation rather than a derivative, since the Jacobi identity for commutators behaves similarly but not identically to the Leibniz product rule. 
We now have the conceptual justification for treating the matrix Laplacian as a second-order derivative operator, 
\begin{equation} \label{matrix_lap}
  \square = \delta_{ab} [X^a, [X^b,\ \cdot\ ]] = g^{ab} \eth_a \eth_b.
\end{equation}
The coefficients $g^{ab}$ on this Laplacian constitute an effective metric tensor. For canonical operators $\{Q^i, P^i\}$, the metric tensor relates to the Poisson structure as 
\begin{equation}\label{eff_metric}
  g^{ab} = \delta_{a'b'} \Theta^{aa'} \Theta^{bb'} = \delta_{a'b'} [X^a, X^{a'}] [X^b, X^{b'}].
\end{equation}
This metric governs the low-energy physics of Yang-Mills matrix models. When trying to extract a classical geometry from a matrix model, it is this metric that will eventually approximate the metric of a classical manifold. For these canonical operators, we simply have the Euclidean metric, $g^{ab} = \delta^{ab}$.

While the metric $g^{ab}$ is easy to calculate when the Poisson structure $\Theta^{ab}$ is known as with canonical operators, this is not always the case for general matrices $\{ X^a \}$. In particular, for any matrices with a non-zero matrix Laplacian, there will be a certain curvature to the underlying manifold, which must be reflected in the metric.

We can recover the Poisson tensor $\Theta^{ab}$ for these general operators $\{ X^a \}$ by defining it at each classical point $x$ as 
\begin{equation}
  \Theta^{ab}(\vech{x}) = -i \langle \vech{x} | [X^a, X^b] | \vech{x} \rangle.
\end{equation}
This construction lets us calculate the effective metric at a point $x$ as the expectation value of the non-commutative metric taken by quasi-coherent states $|\vech{x}\rangle$, 
\begin{align*} 
  g^{ab}(\vech{x}) &= \langle \vech{x} | \delta_{a'b'} \Theta^{aa'} \Theta_{bb'} |\vech{x}\rangle \\ 
  &= \langle \vech{x} | \delta_{a'b'} [X^a, X^{a'}] [X^b, X^{b'}] |\vech{x}\rangle. \numberthis
\end{align*}
In this way, we can characterize the classical geometry associated with any set of matrix operators.

\subsection{Finite representations of matrix geometries via the Jordan-Schwinger construction}
\label{app:jordan-schwinger}

\subsubsection{Non-commutative sphere}

For the matrix representation of a non-commutative sphere $\{rJ_a^{(N)}\}$, we use $N$-dimensional representations of the generators of $SU(2)$ \cite{Madore1991,Hoppe1982,Grosse1996}. 
These operators obey the angular momentum commutation relation, 
\begin{equation}
  [J^{(N)}_i, J^{(N)}_j] = \varepsilon_{ijk} J^{(N)}_k 
\end{equation}
where $\varepsilon_{ijk}$ is the Levi-Civita symbol, i.e.~the totally anti-symmetric tensor of rank three. 

The matrix Laplacian for these operators takes the form $\square X^b=[X^a,[X_a,X^b]]=4r^2 X^b$, indicating a positive eigenvalue and hence uniform positive curvature on the encoded geometry. For the three angular momentum operators of this notional spherical geometry, the resultant quasi-coherent states are the spin-coherent states of the Bloch sphere \cite{Radcliffe1971}.

The $N$-dimensional matrix representations of these $SU(2)$ generators are realized in terms of $j = (N-1) / 2$ and $m_i = i - j$ via raising and lowering operators on an $N$-dimensional basis $|i\rangle$ as 
\begin{equation}
  J_{+} | i \rangle = \sqrt{ \frac{j(j+1) - m_i (m_i + 1)}{2} } |i+1\rangle
\end{equation}
with the $i$ index ranging from 0 to $N-1$ and $J_{-} = J_{+}^\dagger$. The matrix state $ \{J^{(N)}_a\} $ is then defined as 
\begin{align*}
  J_x^{(N)} = \frac{1}{\sqrt{2}}(J_+ + J_-) \quad & \quad J_y^{(N)} = \frac{i}{\sqrt{2}}(J_+ - J_-) \\ 
   J_z^{(N)} = \frac{1}{2} & [J_+, J_-] 
\end{align*}
reproducing the highest-weight representations of $SU(2)$. 

\subsubsection{Non-commutative plane}

For the Moyal-Weyl quantum plane $\{Q^i,P^i\}$, 
the operators are finite representations of canonical position and momentum operators obeying  $[Q^i,P^j]=i\hbar\delta^{ij}\openone$. The Stone-von Neumann theorem \cite{vonNeumann1931,vonNeumann1932,Stone1930,Stone1932}
forbids finite Hermitian representations of the canonical commutation algebra, but approximations of it can be constructed via a Jordan-Schwinger construction using finite-dimensional ladder operators \cite{Jordan1935,Schwinger1952}. 
Ladder operators $a_i$ and $a_j^\dagger$ are operators obeying the commutation relation 
\begin{equation}
  [a_i, a_j^\dagger] = \delta_{ij} \openone.
\end{equation}
The Stone-von Neumann theorem forbids finite representations of these as well, but they can be approximately represented by $N \times N$ matrices $a^{(N)}$ and $a^{(N)\dagger}$, whose entries are given by 
\begin{align*}
  (a^{(N)})_{k,k+1} = \sqrt{k}, && (a^{(N)\dagger})_{k+1,k} = \sqrt{k}
\end{align*}
for $k$ running from $1$ to $N$. The commutation relation for this finite representation is then 
\begin{equation}
  [a_i^{(N)}, a_j^{(N)\dagger}] = \delta_{ij} \text{diag}(1, 1, \dots, 1, -N+1)
\end{equation}
where $a_i^{(N)} = \openone \otimes \openone \otimes \cdots \otimes a^{(N)} \otimes \cdots \otimes \openone$, so that $a^{(N)}$ is embedded in the $i$th position in a larger tensor product of identity matrices, ensuring that $a_i^{(N)}$ and $a_j^{(N)\dagger}$ only commute non-trivially when $i = j$. 

This commutation relation for finite ladder operators does not perfectly reflect the result for ideal ladder operators, but resultant algebraic relations and eigenvalue spectra of operators constructed from these finite ladder operators approaches the desired behavior in the large-$N$ limit. In particular, edge effects appear only at the highest eigenvalues and affect only the maximally excited states of the matrix model. Either way, for matrix models involving finite matrix variables on finite quantum simulator hardware, there is no choice but to encode matrix states in terms of these finite constructions. 

These finite ladder operators immediately provide a finite encoding of canonical position and momentum operators, as 
\begin{align}
  Q_i^{(N)} = \frac{a_i^{(N)} + a_i^{(N)\dagger}}{\sqrt{2}} && P_j^{(N)} = \frac{i (a_j^{(N)} - a_j^{(N)\dagger})}{\sqrt{2}}.
\end{align}
The resultant matrix state obeys the commutation relation 
\begin{equation}
  [Q_i^{(N)}, P_j^{(N)}] = i \delta_{ij} \text{diag}(1, 1, \dots, 1, 1-N)
\end{equation}
producing a finite approximation of the canonical commutation relation for non-relativistic position and momentum operators $\hat{q}_i$ and $\hat{p}_j$.  To the extent that the resultant matrix state obeys the canonical commutation relations, the matrix Laplacian will vanish, ensuring the flatness of the encoded geometry.

\subsubsection{Non-commutative hyperboloid}

For any realistic simulation of a relativistic quantum field, the background spacetime must possess Lorentzian symmetries. To represent the simplest such geometry, namely flat Minkowski spacetime, we need operators associated with a hyperbolic geometry. 
These can be built from the first five generators of the Lie algebra $SO(4,2) = \{ M^{\mu\nu} \}$. 
\begin{equation}
  X^\mu = \{ r M^{\mu 5} \}.
\end{equation} 
These matrices obey the relation $-(X^0)^2+\sum_{i=1}^4(X^i)^2=-R^2\openone$ reproducing the Minkowski signature. This construction dates back to Snyder's early attempts to establish foundations for quantum field theory using appropriately quantized covariant spacetime operators 
\cite{Snyder1947}, and has since been related to the covariant manifolds in general \cite{Steinacker2009} and the FLRW metric in particular \cite{Sperling2019,Steinacker2024}. 

Just as with the spherical geometry, these matrices are also eigenmodes of their matrix Laplacian. 
\begin{equation}
  \square X^\nu = [X^\mu, [X_\mu, X^\nu]] = -4 r^2 X^\nu.
\end{equation}
The magnitude of the eigenvalue indicates the radius of the throat of a hyperboloid embedded in a five-dimensional space, and its negative sign indicates that the manifold has a negative scalar curvature. In this case, the fifth dimension is necessary to ensure that the four physical dimensions are appropriately intertwined. 

The representations of these matrix elements can be defined in terms of a Jordan-Schwinger construction of the generators $M^{ab}$, for which
  \begin{align*}
    M^{ab} = \bar{Z} \Sigma^{ab} Z \quad & \quad\quad  Z = \begin{pmatrix}
      a_1^\dagger & 
      a_2^\dagger &  
      b_1 & 
      b_2 
    \end{pmatrix}^T \numberthis \\ 
    \bar{Z} = Z^\dagger \gamma^0 &= \begin{pmatrix} 
      -a_1, -a_2, b_1^\dagger, b_2^\dagger 
    \end{pmatrix} 
  \end{align*} 
where the ladder operators $a_1$, $a_2$, $b_1$ and $b_2$ obey 
\begin{align}
  [a_i, a_j^\dagger] = \delta_{ij} && [b_i, b_j^\dagger] = \delta_{ij} && \text{ for } i,j = 1,2
\end{align}
and the generators of the Lie algebra $SO(4,2)$ are $\Sigma_{ab}$, defined explicitly as 
\begin{align}
  \Sigma_{\mu\nu} = \frac{1}{4i} [\gamma_\mu, \gamma_\nu] && \Sigma_{\mu 4} = -\frac{i}{2} \gamma_\mu \gamma_4 && \Sigma_{\mu 5} = \frac{1}{2} \gamma_\mu && \Sigma_{45} = \frac{1}{2} \gamma_4 
\end{align}
with $\gamma_\mu$ as the Dirac gamma matrices for $\mu = 0, 1, 2, 3$ and  $\gamma_4 = \gamma_0 \gamma_1 \gamma_2 \gamma_3$. The gamma matrices are the generators of $SO(4,1)$, and obey the anticommutation relation $\{ \gamma_a, \gamma_b \} = -2\eta_{ab}$ for $a, b = 0,1,2,3,4$ and $\eta_{ab} = \text{diag}(-1, 1, 1, 1, 1)$. 
By making use of finite ladder operators $a_i^{(N)}$ and $b_j^{(N)}$, a finite approximation of this hyperboloid algebra can be generated, providing a finite matrix representation of flat Minkowski spacetime.

\subsubsection{Cosmological FLRW metric}
A $k=-1$ FLRW spacetime is encoded by projecting the hyperboloid operators $X^\mu$ to four dimensions ($\mu\in\{0,1,2,3\}$) and introducing conjugate momentum operators, 
\begin{equation} 
  T^\mu=M^{\mu4}/R
\end{equation} 
with $M^{\mu\nu}$ generators of $SO(4,2)$. The resulting matrix Laplacian,
\begin{equation} 
  \square\phi=[T_\mu,[T^\mu,\phi]]=(C^2[SO(4,1)]-C^2[SO(3,1)])\phi, 
\end{equation}
governs the inflationary dynamics. The evaluation of the scale parameter $a(t)$ of this covariant matrix state in the classical large-$N$ limit is somewhat involved \cite{Sperling2019_2,Steinacker2024}, but converges to a $k=-1$ FLRW metric $ds^2 = -dt^2 + a^2(t) d\Sigma^2$, with a scaling parameter $a(t)$ governed by the equations, 
\begin{subequations}
  \begin{align}
    a(t)^2 &= R^2 \sinh \eta \cosh^2\eta \\ 
    dt &= R (\sinh \eta)^{3/2}\,d\eta
  \end{align}
\end{subequations}
where $\eta$ is the cosmological time. As such, at early times, we can approximate $ R \eta^{3/2}\,d\eta = dt $ for $\eta \propto t^{2/5}$, such that $a(t) \propto \eta^{1/2} \propto t^{1/5}$, which recovers the observed age of the universe after inserting the present Hubble rate, without assuming any specific matter content \cite{Sperling2019_2}. At late times, the scaling parameters converges to a more leisurely rate of expansion of $a(t) \approx \frac{3}{2}t$.

A $k=0$ flat FLRW variant exists by choosing null-direction generators satisfying $[T^i,T^j]=0$ and $[T^0,T^i]=(i/R)T^i$, which is the algebra of the Euclidean group $E(3)$. Specifically, the matrix state is defined in terms of the $SO(4,2)$ generators $M^{ab}$ as   
\begin{subequations}
  \begin{align}
  X^a = M^{ac} \alpha_c && a = 0, \dots, 4 \\ 
  T^\mu = M^{\mu d} \beta_d && \mu = 0, \dots, 3 
\end{align}
\end{subequations}
where $\alpha$ and $\beta$ are vectors defined as 
\begin{align}
  \alpha_a = r (0, 0, 0, 0, 0, 1) && \beta_a = \frac{1}{R} (1, 0, 0, 0, 1, 0)
\end{align}
such that $\alpha \cdot \beta = \beta \cdot \beta = 0$ and $\alpha \cdot \alpha = -r^2$. These matrices encode a flat $k = 0$ FLRW metric, in the large-$N$ classical limit \cite{Sperling2019_2}, with the form $ds^2 = dt^2 + a(t)^2 d\Sigma^2$ and a scale parameter given by the equations 
\begin{subequations}
  \begin{align}
    a(t)^2 &= y_0 \\
    dt &= y_0^{1/2}\,dy_0
  \end{align}
\end{subequations}
where $y_0$ is a time-like coordinate. Solving these equations, we get that $t = \frac{2}{3} y_0^{2/3}$ and thus $a(t) \propto t^{1/3}$. These dynamics are different from those of the $k=-1$ covariant matrix state in a manner that better fits observational data in the late-time regime. Nailing down an exact matrix state that corresponds most thoroughly to observational models of cosmology is a crucial future avenue of research, but these analytic solutions indicate the sufficient breadth of the possibility space for such states. 

\subsection{Encoding bosonic fields}
\label{app:fields}

\subsubsection{Scalar fields in matrix models}
Scalar fields can be added to a background $n$-dimensional geometry expressed by $2n$ operators $T^\mu$ by adding in a transverse direction in which the manifold can fluctuate. In particular, we consider a set of $D$ operators such that 
\begin{equation}
    T^a = \begin{cases} 
        T^\mu, & \mu = 1, \dots, 2n \\ 
        \phi^i, & i = 1, \dots, D - 2n.
    \end{cases}
\end{equation}
With the addition of these operators for $D - 2n$ transverse directions, we end up with an action for $D - 2n$ scalar fields, 
\begin{equation}
    S[T^a] = S[T^\mu] 
    + \frac{1}{g^2} \text{Tr}(2[T^\mu, \phi^i][T_\mu, \phi_i] + [\phi^i, \phi^j][ \phi_i, \phi_j]).
\end{equation}
In this action, the $[T^\mu, \phi^i][T_\mu, \phi_i]$ term takes on the role of a kinetic term for scalar fields $\phi^i$ while $[\phi^i, \phi^j][ \phi_i, \phi_j]$ serves as a quartic interaction term between all of the scalar fields.

\subsubsection{Yang-Mills gauge field operators}

To simulate a Yang-Mills gauge field, Hermitian operators that contain a $SU(n)$ symmetry are needed. For example, consider Hermitian operators $T^a$ with a block-diagonal matrix form.
Fluctuations $\mathcal{A}^a$ of the gauge field appear to the extent that the matrices $T^a$ diverge from the matrices of the fixed geometry encoded by $\bar{T}^a$, such that 
\begin{equation}
  T^a = \bar{T}^a + \mathcal{A}^a.
\end{equation}
The matrix field-strength tensor is then 
\begin{equation}
  \mathcal{F}^{ab} = -\Theta^{ab} - \Theta^{aa'} \Theta^{bb'} F_{a'b'}
\end{equation}
where $F_{ab} = \partial_a A_b - \partial_b A_a - i [A_a, A_b] $ is the Yang-Mills gauge field strength tensor for the gauge field potential $A_a = \Theta_{ab}^{-1} \mathcal{A}^b$. The matrix model action in terms of these quantities becomes the familiar action for a $SU(n)$ Yang-Mills gauge theory, 
\begin{equation}
    S = -\frac{1}{g^2} \text{Tr}\ \mathcal{F}^{ab} \mathcal{F}_{ab}.
\end{equation}
The corresponding matrix energy-momentum tensor is 
  \begin{align*}
    \mathcal{T}^{ab} &= \frac{1}{2} ([T^a, T^c] [T^b, T_c] + [T^b, T^c] [T^a, T_c]) \\
    &- \frac{1}{4} \eta^{ab} ([T^c, T^d] [T_c, T_d]) \numberthis
  \end{align*}
with $\eta^{ab}$ chosen for a Lorentzian matrix model, replaced by $\delta^{ab}$ for a Euclidean matrix model. 
The matrix model obeys the following Schwinger-Dyson equation, which also serves as a Ward identity: 
\begin{equation}
  \left\langle [T_a, \mathcal{T}^{ab}] \right\rangle = 0.
\end{equation}
This Ward identity is what preserves the gauge invariance of any non-abelian Yang-Mills field theory represented in the matrix model. In this respect, it takes on the role of the Gauss's law constraints that are typically imposed on lattice gauge theory simulations of a field. 

Note that the $U(n)$ gauge fields encoded in these matrix states are  non-commutative field theories on non-commutative geometries \cite{Seiberg1997}, subject to non-commutative anomalies and their associated UV regularization properties \cite{Szabo2003,ChongSun2001}. The $U(n)$ gauge group can be broken to $SU(n) \times U(1)$ gauge fields by an appropriate spontaneous symmetry-breaking in the matrix state, such as matrix states with intersecting geometries \cite{Grosse2010,Steinacker2018}. 
Alternatively, for the purposes of quantum simulation we can restrict the space of permissible perturbations to an $SU(n)$ basis rather than a $U(n)$ basis, and impose an $SU(n)$-gauge symmetry by hand within the matrix model. This suffices for the purpose of producing a physically relevant simulation, in particular for the quantum circuit setups in which perturbations already have to be specified manually, as detailed in Appendices \ref{globally-controlled-d} and \ref{locally-controlled-e}. 

\subsubsection{Coupling scalar fields and gauge fields}

If one includes operators for directions tranverse to the background geometry $T^\mu$ for operators already possessing $SU(n)$ symmetry, then we have a background geometry given by, 
\begin{equation}
    \bar{T}^a = \begin{cases}
        \bar{X}^\mu, & \mu = 1, \dots, 2n \\ 
        0, & i = 1, \dots, D - 2n 
    \end{cases}
\end{equation}
while the fluctations of this background can be written as 
\begin{align}
    T^a = \bar{T}^a + \mathcal{A}^a, && \mathcal{A}^a = \begin{pmatrix}
        \mathcal{A}^\mu \\ \phi^i 
    \end{pmatrix}.
\end{align}
The commutators of these operators with an arbitrary operator $f$ and with themselves result in covariant derivatives $D_\mu$ and field strength tensors $F_{\mu\nu}$, such that 
\begin{align*}
    [X^\mu, f] &= [\bar{X}^\mu + \mathcal{A}^\mu, f] = i \Theta^{\mu\nu}(\partial_\nu f - i[A_\nu, f]) = i \Theta^{\mu\nu} D_\nu f \\ 
    [X^\mu, X^\nu] &= i\Theta^{\mu\mu'} \Theta_{\nu\nu'} (-\Theta_{\mu'\nu'}^{-1} + F_{\mu'\nu'}). \numberthis
\end{align*}
The action for these operators is thus 
\begin{multline}
    S[T^a] = \frac{1}{g^2} \int_{\mathbb{R}^{2n}} \frac{d^{2n} x}{(2\pi L_\text{NC}^2)^n} \bigg( -\gamma^{\mu\mu'} \gamma^{\nu\nu'} F_{\mu\nu} F_{\mu'\nu'} - \gamma^{\mu\nu} \eta_{\mu\nu} \\ 
    -2 \gamma^{\mu\nu} D_{\mu} \phi^i D_\nu \phi^j \delta_{ij} + [\phi^i, \phi^j][\phi^{i'}, \phi^{j'}] \delta_{ii'} \delta_{jj'} \bigg).
\end{multline}
This describes a non-commutative gauge theory coupled to $D - 2n$ scalar fields $\phi^i$ in the adjoint.

\subsection{Floquet sequence derivation}
\label{app:floquet}

When the fidelity of a unitary operation, i.e.~its Loschmidt echo, is evaluated across a Haar measure of random quantum states, 
its expected value, the probability of each state to remain stationary,
evaluates to 
\begin{equation} 
  \langle f \rangle = \int \mu(\psi) |\langle \psi | U |\psi\rangle|^2 = \frac{N + |\text{Tr}(U)|^2}{N(N+1)} 
\end{equation}
where $N$ is the dimension of the Hilbert space \cite{CollinsSniady2006}. This relation was first discussed as a formula for evaluating average gate fidelities on qudits \cite{Horodecki1999,Nielsen2002}. We offer the proof of this relation in S.I.~\ref{app:haarrelations} via Weingarten calculus. 

The expression for $\langle f \rangle$ has a squared trace $|\text{Tr}(U)|^2$, which does not yet have the form of the $\text{Tr}([X_a, X_b]^2)$ terms in the matrix model action of eq.~(\ref{eq:ym_action}), in which the trace is over a commutator squared. 
To introduce these structures, we encode the commutator of a pair of matrices via a four-part symmetric Floquet sequence. 
Let $ X_a $ and $ X_b $ be Hermitian operators acting as Hamiltonians. 
The corresponding unitary time-evolution operators over time $ t $ are given by $U_a(t) = e^{-i \Delta t X_a}$ and $U_b(t) = e^{-i \Delta t X_b}$. The effective Hamiltonian from the composition of these unitary operations is given by the Baker-Campbell-Hausdorff formula\cite{Hall2013,Baker1905,Campbell1897,Hausdorff1906} as 
\begin{align}
  e^{-i \Delta t X_b} e^{-i \Delta t X_a} = \exp\left( -i\Delta t X_a - i\Delta t X_b + \frac{\Delta t^2}{2} [X_a, X_b] + \mathcal{O}(\Delta t^3) \right).
\end{align}
Conveniently, the commutator of $X_a$ and $X_b$ arises in this expansion at second order in $\Delta t$. We isolate this sequence by a four-part symmetric Floquet sequence in $X_a$ and $X_b$, such that 
\begin{align*}
  \mathcal{F}_2(X_a, X_b; \Delta t) &= e^{i \Delta t X_b} e^{i \Delta t X_a} e^{-i \Delta t X_b} e^{-i \Delta t X_a} \\ 
  &= \exp \left(-i \Delta t^2 (i[X_a, X_b]) + \mathcal{O}(\Delta t^3) \right). \numberthis
\end{align*}
This sequence alternates signs and combines operations in a symmetric pattern that can be visualized as a square path through the operator space generated by $X_a$ and $X_b$, as depicted in Fig.~\ref{fig:fidelities}. 
Commutators are anti-Hermitian, and so we factor out an additional $i$ on the effective Hamiltonian $i[X_a, X_b]$ to ensure that it is a Hermitian operator. In effect, we apply a dynamical decoupling operation \cite{Viola1999,Zanardi1999} and make use of the remaining error term. 

If we apply the unitary operation $\mathcal{F}_2(X_a, X_b; \Delta t)$ to a Haar-random ensemble, the expected fidelity $\langle f_{ab} \rangle$ will depend on the value of $|\text{Tr}(\mathcal{F}_2(X_a, X_b; \Delta t))|^2 = |\text{Tr}(\exp \left( -\Delta t^2[X_a, X_b] \right))|^2$. For sufficiently small values of $\Delta t$, this trace can be expanded as 
\begin{align*}
  |\text{Tr}(\mathcal{F}_2(X_a, X_b; \Delta t))|^2 &= |\text{Tr}( \openone - i \Delta t^2(i[X_a, X_b]) \\ 
  &\quad\quad - \Delta t^4(i[X_a, X_b])^2 + \mathcal{O}(\Delta t^6) )|^2 \\  
  &= |N - \text{Tr}\left( i \Delta t^4[X_a, X_b]^2  \right) + \mathcal{O}(\Delta t^6)|^2 \\ 
  &= N^2 - 2 \Delta t^4 N \text{Tr}\left(i [X_a, X_b]^2  \right) + \mathcal{O}(\Delta t^8) \\ 
  &= N^2 - 2 \Delta t^4 N ||i[X_a, X_b]||_F^2 + \mathcal{O}(\Delta t^8). \numberthis
\end{align*}
The term $\text{Tr}(-\Delta t^2[X_a, X_b])$ disappears in the first line by the cyclicity of the trace. Dropping the trailing-order terms and plugging this into the Haar measure expression, we find that 
\begin{align*} \label{eq:frob-haar-si}
  \langle f_{ab} \rangle &= \int d\mu(\psi) |\langle \psi | \mathcal{F}_2(X_a, X_b; \Delta t) |\psi\rangle|^2 \\ 
  &\simeq \frac{N + N^2 - 2 \Delta t^4 N ||i [X_a, X_b]||_F^2}{N(N+1)} \\ 
  &= 1 - \frac{2 \Delta t^4}{N+1} ||i [X_a, X_b]||_F^2. \numberthis
\end{align*}
For this approximation to hold, the higher-order terms must vanish as $\Delta t^6\,\text{Tr}([X_a, [X_a, X_b]]^2) \ll 1 $. 
With a Trotterized Floquet sequence \cite{Suzuki1990} of $\mathcal{F}_2$ operations across all pairs of $X_a$ and $X_b$ drawn from the matrix state $\{ X_a \}$ of the model, the overall fidelity is 
\begin{align*}
  \langle f \rangle &=  \int d\mu(\psi) |\langle \psi |\prod_{a,b} \mathcal{F}_2(X_a, X_b; \Delta t) |\psi\rangle|^2 \\ 
  &\simeq 1 - \frac{2 \Delta t^4}{N+1} \sum_{a,b} ||i [X_a, X_b]||_F^2.  \numberthis
\end{align*}
Comparing this result to the expression for the weights of the Euclidean path integral in eq. (\ref{eq:frob-partition}), we see that their leading-order behaviors are identical under the identification $ \frac{1}{g^2} = \frac{2\Delta t^4}{N+1} $. 
The essential relation for the quantum simulation of the Euclidean path integral comes into focus after the appropriate choice of $\Delta t$ and $N$, giving us the central result presented in equation (\ref{eq:fid-relation}), 
\begin{equation}
  e^{-S[X]} \simeq \int d\mu(\psi) |\langle \psi |\prod_{a,b} \mathcal{F}_2(X_a, X_b; \Delta t) |\psi\rangle|^2.
\end{equation}
Every measurement of the expected fidelity of a Haar-random state under a Floquet sequence is therefore a direct measurement of the Boltzmann weight assigned to that matrix configuration in the path integral.

Extending the Floquet sequence across an entire set of $n$ matrices $\{X_a\}$ can be naively achieved by applying Floquet sequences $\mathcal{F}_2(X_a, X_b; \Delta t)$ for every pair of operators by hand and then Trotterizing these together for a total of $4 (n^2 - n) / 2 $ operations. Alternatively, the same result can be generated with quadratically fewer operations by applying rising and falling sequences of unitaries, introduced in Appendix \ref{floquet-derivation-c} as 
\begin{align*}
  \mathcal{F}_2(\{X_a\}; \Delta t) &= \left( \prod_a e^{i \Delta t X_a} \right) \left( \prod_b e^{-i \Delta t X_b} \right) \numberthis \\ 
  &= \exp\left( - i \Delta t^2 \sum_{a,b} i [X_a, X_b]  + \mathcal{O}(\Delta t^3)\right)
\end{align*} 
With this, the Floquet sequence for generating commutators of operators is generalized to arbitrarily large $n$-element matrix states at the cost of merely $n$ operations. In effect, every unitary operation is applied in periodic steps with a discrete phase offset. This immediately suggests a generalization of this operation to continuous periodically-driven Hamiltonians, with the Magnus expansion taking the place of the BCH formula.  

\subsection{Equivalent dynamics from generalized periodic driving}
\label{app:periodic-drive}

Effective Hamiltonians with leading-order commutator terms are not unique to Floquet sequences such as $\mathcal{F}_2$, as they arise ubiqiutously in Floquet theory studies of periodically driven quantum systems \cite{Bukov2015,Eckardt2017}. 
In particular, any periodic Hamiltonian alternating between two operators generates an effective Hamiltonian whose leading-order contribution is their commutator.

We start by evaluating the Magnus expansion \cite{Magnus1954,Blanes2009} of a periodic Hamiltonian presented as 
\begin{equation}
  H(t) = \sin(\omega t) X_a + \cos(\omega t) X_b.
\end{equation}
This periodic term could be an isolated term or a Fourier component of a generalized periodic Hamiltonian. Its first-order and second-order Magnus terms over a period $\Delta t = \frac{2\pi}{\omega}$ are 
\begin{subequations}
  \begin{align*}
  \Omega_1 &= \int_0^{\Delta t}  dt_1\ H(t_1) \\ 
  &= \int_0^{\Delta t}  dt_1\ \sin(\omega t_1) X_a + \cos(\omega t_1) X_b \\ 
  &= 0 \numberthis \\ 
  \Omega_2 &= \frac{1}{2} \int_0^{\Delta t}  dt_1 \int_0^{t_1} dt_2\ [H(t_1), H(t_2)] \\ 
  &= \frac{1}{2} \int_0^{\Delta t}  dt_1 \int_0^{t_1} dt_2\ (\sin(\omega t_1) \cos(\omega t_2) \\ 
  &\quad\quad\quad\quad\quad\quad\quad\quad\quad\quad - \cos(\omega t_1) \sin(\omega t_2)) [X_a, X_b] \\ 
  &= \frac{1}{2}  [X_a, X_b] \int_0^{\Delta t}  dt_1 \int_0^{t_1} dt_2\ \sin(\omega (t_1 - t_2)) \\ 
  &= -\frac{\pi}{\omega^2}  [X_a, X_b]. \numberthis 
\end{align*}
\end{subequations}
At high frequencies, the $\Omega_2$ term is the dominant component of the Hamiltonian. Over each period $\Delta t$, the effective Hamiltonian is thus 
\begin{equation}
  H_\text{eff} = \frac{\Omega_2}{\Delta t} = -\frac{1}{2\omega}  [X_a, X_b]. \numberthis
\end{equation}
As a result, Floquet driving for these simulations does not need to be engineered with exact sequences of successive $X_a$ and $X_b$ operators, but can instead arise from any periodic driving between operators. 

For arbitrary matrix states $ \{ X_a \} $, this emergence of commutators from periodic driving occurs for any operators applied with amplitude modulation at varying phases, with unitary evolution of the form 
\begin{align*}
  \mathcal{F}_2^\omega(\{X_a\}; \Delta t) &= \exp \left( -i \Delta t \left( \sum_{a} \cos(\omega (t + \phi_a)) X_a \right) \right) \\ 
  &= \exp \left( -i \Delta t \frac{2\pi}{\omega^2(n-1)} \sum_{a,b} [X_a, X_b] + \mathcal{O}\left(\frac{1}{\omega^3}\right) \right)
\end{align*}
where the offset phases $\phi_a$ are distributed across the unit circle so as to induce commutators between the operators at the alternating peaks of the amplitude modulation. 
When these effective Hamiltonians with sums of pairs of commutators are applied to Haar-random quantum ensembles, the resulting expected fidelities will reflect the matrix model path integral just as before with the discrete Floquet sequence.

\subsection{Haar-state identities and Weingarten calculus}
\label{app:haarrelations}

We wish to prove the identity
\begin{equation} 
  \int d\mu(\psi) |\langle \psi | U |\psi\rangle|^2 = \frac{N + |\text{Tr}(U)|^2}{N(N+1)} 
\end{equation}
using Weingarten calculus, where $|\psi\rangle$ is a pure state drawn from the Haar measure, $U$ is a unitary operation, and $N$ is the size of the Hilbert space.

Consider the integrand $|\langle \psi|U|\psi\rangle|^2$. Since all pure states are related by unitary operations, we can rewrite this as
\begin{equation}
  \int d\mu(\psi)|\langle \psi|U|\psi\rangle|^2=\int dV|\langle 0|V^\dagger UV|0\rangle|^2
\end{equation}
where $V$ is a unitary operation. This trades the Fubini-Study integral over all states $|\psi\rangle$ for a Haar integral over unitary operations $V$ with reference state $|0\rangle$. Fixing a basis $\{e_i\}$ with $|0\rangle=e_1$, we have
\begin{subequations}
  \begin{align}
    \langle 0|V^\dagger UV|0\rangle &=\sum_{i,j}\overline{V_{i1}} \, U_{ij}V_{j1}\\
    |\langle 0|V^\dagger UV|0\rangle|^2 &=\sum_{i,j,k,l}U_{ij}\overline{U_{kl}}V_{j1}V_{l1}\overline{V_{i1}}\,\overline{V_{k1}}.
  \end{align}
\end{subequations}
Plugging back into the Haar integral gives
\begin{align*}
  &\int dV \sum_{i,j,k,l}U_{ij}\overline{U_{kl}}V_{j1}V_{l1}\overline{V_{i1}}\overline{V_k1}\\
  &=\sum_{i,j,k,l}U_{ij}\overline{U_{kl}}\int dV V_{j1}V_{l1}\overline{V_{i1}}\,\overline{V_{k1}}
  \label{D6}. \numberthis
\end{align*}
The general $U(N)$ Weingarten identity \cite{CollinsSniady2006} is
\begin{align*}
  &\int_{U(N)} U_{i_1 j_1} \cdots U_{i_d j_d} \,
  \overline{U_{k_1 l_1}} \cdots \overline{U_{k_d l_d}} \, dU\\
  &=
  \sum_{\sigma, \Delta t \in S_d}
  \prod_{s=1}^{d} \delta_{i_s,\, k_{\sigma(s)}}\,
  \delta_{j_s,\, l_{\tau(s)}}
  \mathrm{Wg}(\sigma^{-1}\tau,\, N). \numberthis
\end{align*}
Intuitively, the Haar measure of the integral over $U(N)$ forces each $U$ index to contract with a $\overline{U}$ index. This results in a sum over permutations $(\sigma,\tau)\in S_d\times S_d$ with the Kronecker deltas enforcing the contractions. The Weingarten function $Wg(\sigma^{-1}\tau,N)$ contains all the necessary coefficients for the permutations. For a degree-2 integral, the identity is
\begin{align*}
  &\int_{U(N)} U_{a_1 b_1} U_{a_2 b_2} \overline{U_{c_1 d_1}} \, \overline{U_{c_2 d_2}} \, dU \\
  &= \sum_{\sigma, \Delta t \in S_2} \prod_{s=1}^{2} \delta_{a_s,\, c_{\sigma(s)}} \, \delta_{b_s,\, d_{\tau(s)}} \operatorname{Wg}(\sigma^{-1}\tau,\, N). \numberthis
\end{align*}
Using this identity, the Haar integral over $V$ can be re-expressed in terms of a Weingarten function.
\begin{equation}
  \int dV\,\overline{V_{i1}}\overline{V_{k1}}V_{j1}V_{l1}=\sum_{\sigma,\tau\in S_2}\delta_{j,c_\sigma(1)}\delta_{k,c_\sigma(2)}\mathrm{Wg}(\sigma^{-1}\tau,\, N).
\end{equation}
The group $S_2=\{e,(12)\}$ has two elements, so the sum over $S_2\times S_2$ contains four terms. Evaluating the sum gives
\begin{equation}
  \int dV\,\overline{V_{i1}}\overline{V_k1}V_{j1}V_{l1}=(\delta_{j,i}\delta_{k,l}+\delta_{j,l}\delta_{k,i})(\mathrm{Wg}(e,N)+\mathrm{Wg}((12),N)).
\end{equation}
The Weingarten function for $U(N)$ at degree $d$ is solved and given by
\begin{equation}
  \text{Wg}(\pi,N) = \sum_{\lambda \vdash d} \frac{\chi^{\lambda}(\pi)}{\prod_{\square \in \lambda}(N + c(\square))} \cdot \frac{\dim V^{\lambda}}{d!}
\end{equation}
where $\pi\in \text{S}(d)$, $\lambda$ refers to the Young diagrams of $\text{S}(d)$, $\chi^\lambda$ is the character of $V^\lambda$, and $c(\square)$ picks out the content of a Young diagram. For $\text{S}(2)$, there are two irreducible represenations corresponding to Young diagrams $\lambda=[2]$ and $\lambda=[1,1]$. To evaluate the denominator of the Weingarten functions, the Young diagrams have contents $\{0,1\}$ and $\{0,-1\}$, respectively. Thus, the denominators are
\begin{equation}
  ~~~~~~\prod_{\square\in\lbrack2\rbrack}=N(N+1)~~~~\prod_{\square\in\lbrack1,1\rbrack}=N(N-1).
\end{equation}
As for the characters, $\lambda=\lbrack2\rbrack$ is the trivial representation, and so $\chi^{\lbrack2\rbrack}(e)=\chi^{\lbrack2\rbrack}((12))=1$. The other representation $\lambda=\lbrack1,1\rbrack$ is the sign representation, so $\chi^{\lbrack 1,1\rbrack}(e)=1$ and 
$\chi^{\lbrack1,1\rbrack}((12))=-1$. Finally, $\dim V^\lambda=1$ for both representations. We want to calculate $\text{Wg}(e,N)$ and $\text{Wg}((12),N)$. Using the above calculations, these evaluate to
\begin{align*}
  \text{Wg}(e,N)&=\frac{1}{2N(N+1)}+\frac{1}{2N(N-1)}\\
  &=\frac{1}{N^2-1} \numberthis
\end{align*}
\begin{align*}
  \text{Wg}((12),N)&=\frac{1}{2N(N+1)}-\frac{1}{2N(N-1)}\\
  &=-\frac{1}{N(N^2-1)}. \numberthis
\end{align*}
Summing these gives
\begin{equation}
  \text{Wg}(e,N)+\text{Wg}((12),N)=\frac{1}{N(N+1)}.
\end{equation}
Thus,
\begin{equation}
  \int dV \overline{V_{i1}}\overline{V_{k1}}V_{j1}V_{l1}=\frac{1}{N(N+1)}(\delta_{j,i}\delta_{k,l}+\delta_{j,l}\delta_{k,i}).
\end{equation}
Plugging this back into \eqref{D6} gives
\begin{align*}
  &\frac{1}{N(N+1)}\sum_{i,j,k,l}U_{ij}\overline{U_{kl}}(\delta_{j,i}\delta_{k,l}+\delta_{j,l}\delta_{k,i})\\
  &=\frac{1}{N(N+1)}\left(\sum_{i,k}U_{ii}\overline{U_{kk}}\right)+\left(\sum_{i,j}U_{ij}\overline{U_{ij}}\right)\\
  &=\frac{\lvert\text{Tr}(U)\rvert^2+N}{N(N+1)}. \numberthis
\end{align*}
Thus, we have that
\begin{equation}
  \int d\mu(\psi)|\langle \psi|U|\psi\rangle|^2=\frac{N+\lvert\text{Tr}(U)\rvert^2}{N(N+1)}.
\end{equation}
This proves the desired identity, and provides the expression for evaluating the average Loschmidt echo of Haar-random states under a given unitary operation. 

\end{document}